\documentclass[sigconf]{acmart}
\AtBeginDocument{%
 \providecommand\BibTeX{{%
 \normalfont B\kern-0.5em{\scshape i\kern-0.25em b}\kern-0.8em\TeX}}}
\copyrightyear{2020}
\acmYear{2020}
\setcopyright{acmlicensed}\acmConference[ACSAC 2020]{Annual Computer Security Applications Conference}{December 7--11, 2020}{Austin, USA} \acmBooktitle{Annual Computer Security Applications Conference (ACSAC 2020), December 7--11, 2020, Austin, USA}
\acmPrice{15.00}
\acmDOI{10.1145/3427228.3427287} 
\acmISBN{978-1-4503-8858-0/20/12}
\usepackage{makecell}
\usepackage{enumitem}
\usepackage{acmart-taps}
\usepackage{tablefootnote}

\usepackage{wasysym}

\usepackage{subfigure}

\newcommand{\quotesd}[1]{``#1''}

\newcommand{\subhead}[1]{\vspace*{1pt}\noindent{\textbf{#1.}}} %\usepackage{pifont}
\newcommand{\formatSpecial}[1]{\texttt{#1}}

\newcommand{\testPeriod}{Mar.\ 2019 to May 2020} 
 
\newcommand{\meetcircle}{Circle}
\newcommand{\familytime}{FamilyTime}
\newcommand{\famisafe}{FamiSafe}
\newcommand{\findmykids}{FindMyKids}
\newcommand{\kidsplace}{KidsPlace}
\newcommand{\KidControl}{KidControl}
\newcommand{\life}{Life360}
\newcommand{\mobilefence}{MobileFence}
\newcommand{\qustodio}{Qustodio}
\newcommand{\screentime}{ScreenTime}
\newcommand{\secureteen}{SecureTeen}
\newcommand{\mmguadian}{MMGuardian}
\newcommand{\kidoz}{Kidoz}
\newcommand{\apiEp}{API endpoint}
\newcommand{\apiReq}{API request}
\newcommand{\kidswifi}{KidsWifi}
\newcommand{\blocksi}{Blocksi}
\newcommand{\networkdevices}{network devices}
\newcommand{\Networkdevices}{Network devices}
\newcommand{\networkdevice}{network device}
\newcommand{\androidapps}{Android apps}

\newcommand{\numofandroidappsDA}{29}
\newcommand{\numofandroidsol}{13}

\newcommand{\selecteddownloaded}{158}
\newcommand{\totalselected}{168}
\newcommand{\totalevaluated}{153}
\newcommand{\androidsol}{Android solution}
\newcommand{\androidsols}{Android solutions}
\newcommand{\app}{app}

\newcommand{\androidapp}{Android app}
\newcommand{\windowsapps}{Windows applications}
\newcommand{\windowsapp}{Windows application}

\newcommand{\extension}{Chrome extension}
\newcommand{\extensions}{Chrome extensions}
\newcommand{\pcsproducts}{parental control tools}
\newcommand{\pcsproduct}{parental control tool}
\newcommand{\Pcsproducts}{Parental control tools}
\newcommand{\purechildrenapps}{children apps}
\newcommand{\mixedchildparentenapps}{shared apps}

\newcommand{\purechildrenapp}{children app}

\newcommand{\parentapps}{parent apps}

\newcommand{\HPurechildrenapps}{\makecell[c]{Children\\apps}}
\newcommand{\HMixedchildparentenapps}{\makecell[c]{Shared\\apps}}
\newcommand{\HParentapps}{\makecell[c]{Parent\\apps}}
\newcommand{\SDK}{SDK}
\newcommand{\SDKs}{SDKs}
\newcommand{\domain}{SDK} 
\newcommand{\domains}{SDKs}

\newcommand{\cxmark}{\textcolor{black}{\ding{51}}\textsuperscript{\textcolor{red}{\kern-0.55em\ding{55}}}}

\newcommand{\partiallycomply}{\textcolor{black}{\ding{51}}\textsuperscript{\textcolor{red}{\kern-0.55em\ding{55}}} }

\usepackage{float}
\usepackage{multirow}

\usepackage{tikz}

\begin{document}

\title[Betrayed by the Guardian]{Betrayed by the Guardian:\\ Security and Privacy Risks of Parental Control Solutions}

\author{Suzan Ali}
\email{a\_suzan@ciise.concordia.ca}
\affiliation{%
 \institution{Concordia University}
 \city{Montreal}
 \state{Quebec}
 \country{Canada}
}

\author{Mounir Elgharabawy}
\email{m\_elghar@encs.concordia.ca}
\affiliation{
 \institution{Concordia University}
 \city{Montreal}
 \state{Quebec}
 \country{Canada}
}

\author{Quentin Duchaussoy}
\email{q\_duchau@encs.concordia.ca}
\affiliation{
\institution{Concordia University}
\city{Montreal}
\state{Quebec}
\country{Canada}
}

\author{Mohammad Mannan}
\email{m.mannan@concordia.ca}
\affiliation{
\institution{Concordia University}
 \city{Montreal}
 \state{Quebec}
 \country{Canada}
}

 \author{Amr Youssef}
 \email{youssef@ciise.concordia.ca}
 \affiliation{
 \institution{Concordia University}
 \city{Montreal}
 \state{Quebec}
 \country{Canada}
}

\renewcommand{\shortauthors}{S.\ Ali et al.}

\begin{abstract}
For parents of young children and adolescents, the digital age has introduced many new challenges, including excessive screen time, inappropriate online content, cyber predators, and cyberbullying. To address these challenges, many parents rely on numerous parental control solutions on different platforms, including parental control network devices (e.g., WiFi routers) and software applications on mobile devices and laptops. While these parental control solutions may help digital parenting, they may also introduce serious security and privacy risks to children and parents, due to their elevated privileges and having access to a significant amount of privacy-sensitive data. In this paper, we present an experimental framework for systematically evaluating security and privacy issues in parental control software and hardware solutions. Using the developed framework, we provide the first comprehensive study of {\pcsproducts} on multiple platforms including \networkdevices{}, \windowsapps{}, \extensions{} and \androidapps{}. Our analysis uncovers pervasive security and privacy issues that can lead to leakage of private information, and/or allow an adversary to fully control the parental control solution, and thereby may directly aid cyberbullying and cyber predators. 
\end{abstract}




\maketitle

\section{Introduction}

Many of today’s children cannot imagine their daily lives without internet.
A recent survey~\cite{Sellcellreport} 
shows that 42\% of US children (4--14 years) spend over 30 hours a week on their phones;
nearly 70\% of parents think that such use has a positive effect on their children’s development~\cite{Sellcellreport}.
While the web could be an excellent environment for learning and socializing, there is also a plethora of online content that can be seriously damaging to children. In addition, children are by nature vulnerable to online exploitation and other risk effects of online social networking, including cyber-bullying and even cyber-crimes (see e.g.,~\cite{risks-survey, risks-thesis}); the current COVID-19 pandemic has only increased these risks (see e.g.,~\cite{covid-un}).

To provide a safe, controlled internet experience, many parents and school administrators rely on parental control solutions that are easily accessible either for free, or for a relatively cheap price. From recent surveys in the US, some forms of parental control apps/services are used by 26--39\% of parents~\cite{csm-usage, pew-usage}, indicating a growing adoption of these solutions. Such solutions are also recommended by government agencies, e.g., US FTC~\cite{reco-ftc} and UK Council for Child Internet Safety (UKCCIS)~\cite{reco-uk}, despite their limited effectiveness (cf.\ EU commissioned benchmark at: \href{https://sipbench.eu}{sipbench.eu}), and questionable morality since they, arguably, can act as surveillance tools~\cite{wisniewski2017parental}. This ethical/moral debate is outside the scope of our work.\looseness=-1

On the other hand, over the past few years, many attacks targeted parental control solutions, exposing monitored children's data, sometimes at a large scale~\cite{TeenSafedatabreachnews,Familyorbitdatabreachnews}. Aside from endangering children's safety (online and in the real-world), such leaked children's personal data may be sold by criminals (cf.~\cite{child-pii-sell}). 
Recent reports also revealed several security and privacy issues in the analyzed parental control solutions~\cite{faesslersafer1,Circledatabreachnews,feal2019angel}. However, such analysis was limited to the privacy of Android apps, and only one network device, even though popular parental control solutions are used across different platforms: mobile and desktop OSes, web extensions, and network devices. Note that, unlike other vulnerable products (e.g., buggy gaming apps~\cite{game-buggy}), or non-complaint products (e.g., Android apps for children~\cite{reyes2018won}), which can be removed when such concerns are known, parental control solutions are deemed \emph{essential} by many parents and schools, and thus are not expected to be removed due to the lack of better alternatives.\looseness=1

We undertake the first comprehensive study to analyze different types of parental control hardware and software solutions. We design a set of security and privacy tests, and systematically analyze popular representative parental control solutions available in \networkdevices{}, Windows and Android OSes, and Chrome extensions.
While developing our comprehensive analysis framework for solutions in multiple platforms, we faced several challenges. Most parental control solutions implement various techniques that hinder traffic analysis (e.g., VPNs, SSLpinning and custom protocols). The use of proprietary firmware and code obfuscation techniques also poses challenges for static analysis. Understanding long-term behaviors of these solutions by running them for hours/days in a realistic way (e.g., triggering all their features), is also time consuming (compared to simple, automated UI fuzzing). 

\subhead{Contributions}
Our contributions can be summarized as follows. (i) We developed an experimental framework for systematically evaluating security and privacy issues in parental control software and hardware solutions. (ii) We utilized this framework to conduct the first comprehensive study of {\pcsproducts} on multiple platforms, including 8 \networkdevices{}, 8 \windowsapps{}, 10 \extensions{}, and \numofandroidappsDA{} \androidapps{} representing \numofandroidsol{} \androidsols{} grouped by vendor.\footnote{An Android solution is typically composed of a child app, a parent app, and an online parental dashboard. We consider an \androidsol{} vulnerable if any of its component is vulnerable.} The in-depth analysis aims to inspect the apps' web traffic for personally identifiable information (PII) leakage, insecure \apiEp{}s authentication, potential vulnerabilities, and the presence of third-parties and known trackers. 
(iii) Our analysis reveals 135 vulnerabilities among the solutions tested and highlights that the majority of solutions broadly fail to adequately preserve the security and privacy of their users---both children and parents.

\subhead{Notable findings and disclosure}
Our notable findings include:

\begin{itemize}[noitemsep,topsep=0pt,parsep=0pt,partopsep=0pt,leftmargin=5mm]

\item The \blocksi{} parental control router allows remote command injections, enabling an attacker with a parent's email address to eavesdrop/modify the home network's traffic, or use the device in a botnet (cf.\ Mirai~\cite{mirai}). \blocksi{}'s firmware update mechanism is also completely vulnerable to a network attacker.

\item 8/\numofandroidsol{} \androidsols{} and 4/8 \networkdevices{} do not properly authenticate their server \apiEp{}s, allowing illegitimate access to view/modify server-stored children/parents data.

\item 5/\numofandroidsol{} \androidsols{} allow an attacker to easily compromise the parent account at the server-end, enabling full account control to the child device (e.g., install/remove apps, allow/block phone calls and internet connections). 

\item 7/\numofandroidsol{} \androidsols{} transmit PII via HTTP (e.g.,  kidSAFE~\cite{kidozsafeseal} certified \kidoz{} sends account credentials via HTTP). 

\item Among the \pcsproducts{} with a web interface, 9/\numofandroidsol{} \androidsols{}, 4/8 \networkdevices{}, and 3/8 \windowsapps{} are vulnerable to  SSLStrip attacks (cf.~\cite{hsts-ndss19, li2017hsts}), a man-in-the-middle (MITM) attack, due to the lack of HSTS.

\item 2/8 \windowsapps{} utilize a TLS proxy that degrades connection security, by accepting certificates and ciphers that are rejected by modern browsers. Another Windows application (Kidswatch) completely lacks HTTPS, and communicates with the backend server via HTTP. 

\item 2/10 \extensions{} and 4/\numofandroidsol{} \androidsols{} transmit the full URLs from the browser to their server, possibly leaking sensitive (session) information. 
\end{itemize}

As part of responsible disclosure, we contacted the developers of the solutions we analyzed, and shared our findings, including proof-of-concept scenarios and possible fixes. Two months after disclosure, only ten 
companies
responded, seven custom and three automatic replies. 
Blocksi, KoalaSafe, \mmguadian{}, \kidsplace, \famisafe{}, and \familytime{} responded that they are investigating the issues. \kidoz{} responded that some of our reported issues are on their fixing backlog, and acknowledged the new vulnerabilities. Other vendors either sent automatic/ambiguous response, or no response at all. 
Notable changes after the disclosure: \mmguadian{} deprecated their custom browser; FamiSafe fixed the Firebase database security issue; and \familytime{} enabled HSTS on their server.
 
\section{Related Work}

In this section, we first list a few example cases from real-world data breaches involving \pcsproducts{}, and then summarize related academic studies (mostly privacy analyses of Android apps).

Over the past years, several \pcsproducts{} have made the news for
security and privacy breaches. 
The teen-monitoring app TeenSafe leaked thousands of children's Apple IDs, email addresses and passwords~\cite{TeenSafedatabreachnews}.
Family Orbit exposed nearly 281 GB of children data from an unsecured cloud server~\cite{Familyorbitdatabreachnews}. In 2019, a privacy flaw in Kaspersky anti-virus and parental control application was found~\cite{kasperskyScript}. This application included a script to perform content checking on each page intercepted by a TLS proxy. However, some unique IDs were also included in the process, allowing the website to track the user. In 2010, EchoMetrix settled US FTC charges for collecting and selling children's information to third-parties through their parental control software~\cite{EchoMetrix}.

Between 2015 and 2017, researchers from the Citizen Lab (\href{https://citizenlab.ca/}{citizenlab.ca}), Cure53 (\href{https://cure53.de/}{cure53.de}), and OpenNet Korea (\href{http://opennetkorea.org/en/wp/}{opennetkorea.org}) published a series of technical audits of three popular Korean parenting apps mandated by the Korean government, Smart Sheriff, Cyber Security Zone and Smart Dream~\cite{faesslersafer1}. The security audits found serious security and privacy issues in the three parental control Android apps. For example, Smart Sheriff failed to adequately encrypt PII either on storage or in transit. Smart Dream allowed unauthorized access to children's messages and search history.

Feal et al.~\cite{feal2019angel} studied 46 parental control Android apps for data collection and data sharing practices, and the completeness and correctness of their privacy policies. They used the Lumen Android app (see \url{https://haystack.mobi/}) for their analysis, which is unable to analyze target apps with VPN or certificate pinning. Parental apps and dashboards are also excluded. Our analysis framework has no such limitations, and consequently we are able to identify
new critical security issues (e.g., leakage of plaintext authentication information), even among the apps analyzed by Feal et al. 

Reyes et al.~\cite{reyes2018won} analyzed children Android apps for COPPA compliance.  
Out of 5855 apps, the majority of the analyzed apps were found to potentially violate COPPA, and 19\% were found to send PII in their network traces. 
Wisniewski et al.~\cite{wisniewski2017parental} evaluated 42 features in 75 parental control \androidapps{}, showing that most apps value 
control over self-regulation strategies,
and boast the use of privacy invasive techniques. Marsh~\cite{marsh2018examination} 
measured the effectiveness and usability of two parental control apps.

Web extensions have been subjected to security evaluation for over a decade (see e.g.,~\cite{starovextended, chenmystique}), but no past studies focused on parental control extensions.
Windows parental control applications have been only studied for the security of their TLS proxies~\cite{de2016killed}. 
Similarly, parental control network devices remained unexplored, except the Disney Circle, analyzed by Cisco Talos in 2017, and found to have 23 different security vulnerabilities~\cite{Circledatabreachnews}. 
Among other devices, we also analyzed Circle, but used a newer version released in 2019. 

In contrast to previous work, we conduct a comprehensive, systematic study of security and privacy threats in parental control solutions across multiple platforms: mobile (Android), desktop (Windows), web browser (Chrome extensions) and stand-alone network devices, as popular solutions are available in all these platforms. 
Our analysis therefore sheds light on the broader picture of security and privacy risks of \pcsproducts{}. Compared to existing \androidapp{} studies, our framework is more in-depth (e.g., monitoring the apps from the OS instead of the application level), and inclusive (e.g., analyze apps with VPNs and key pinning). 

\section{Background and Threat Model}

We use the term ``\pcsproducts{}'' to cover different types of parental solutions: \networkdevices{}, \androidapps{}, \extensions{} and \windowsapps{}.
Personally identifiable information (PII) refers to any information related to the user as defined by the US FTC and Office of the Privacy Commissioner of Canada. 
Any entity that is not directly related to a parental control solution, is labelled as a third-party; this includes but is not limited to trackers and advertisers.
In what follows, we briefly discuss some common techniques used by \pcsproducts{}, define our threat model, and list the vulnerabilities that we test against each solution.

\subsection{Monitoring Techniques}\label{sec:montoring-tech}
\Pcsproducts{} generally allow the parent to remotely control the child device, perform web filtering, and monitor activities on social media. We derive the following monitoring techniques from product documentation, our observations from installation procedure and use/analysis of these solutions. These techniques vary significantly across platforms, and are grouped here as such. 

\subhead{\Networkdevices{}}
Being network-based, parental control devices can monitor network traffic but cannot inspect the content of encrypted traffic. The devices analyzed act as a man-in-the-middle between the client device and the internet router by using one of two techniques: performing Address Resolution Protocol (ARP) spoofing, or creating a separate access point. ARP spoofing enables the network device to impersonate the internet router on the local network. The device achieves that by sending forged ARP packets that bind the router's IP with the network device's MAC address. As a result, all the local network traffic is routed through the device before going to the internet router. Alternatively, the network device may create an explicit access point exclusively for children to enforce parental control filtering on all devices connected to it.

\subhead{\androidapps{}}
\androidapps{} rely on several Android-specific mechanisms, including the following (see Table~\ref{table:moniteringtechniques} in Appendix for per \androidsol{} capabilities). 
(1) Device administration~\cite{deviceadmin,shan2019device} provides several administrative features at the system level, including: device lock, factory reset, certificate installation, and device storage encryption. 
(2) Mobile device management (MDM~\cite{madden2013enterprise}) enables additional control and monitoring features, designed for businesses to fully control/deploy devices in an enterprise setting.\footnote{Note that, MDM features may be just too powerful, and may enable dangerous remote control operations including device wipe. Apple has removed several popular parental control apps from App Store due to their use of such highly invasive features (\url{https://www.apple.com/ca/newsroom/2019/04/the-facts-about-parental-control-apps/}). In contrast, Google Play apparently still allows these features in parental apps.} 
(3) Android accessibility service~\cite{accessabilityservice,shan2019device} enables apps to perform several functions including 
monitoring user actions by receiving notifications when the user interacts with an app, capturing and retrieving window content, logging keystrokes, and controlling website content by injecting JavaScript code into visited web pages. 
(4) Notification access enables \androidapps{} to read or dismiss all notifications displayed in the status bar; notifications may include personal information such as contact names and messages. 
(5) Android VPN, custom browsers, and third-party domain classifiers (e.g., Komodia.com~\cite{komodia}), which are used to filter web content.
(6) Facebook~\cite{FacebookLogin} and YouTube OAuth~\cite{youtubeoAuth} features, which are used to monitor the child's activities on Facebook (e.g., posts and photos), and YouTube (e.g., playlists and comments). 
(7) Miscellaneous techniques including: having browser history and bookmarks permission, using custom browsers, or TLS interceptions via Android VPN.\looseness=-1

\subhead{\windowsapps{}}
As opposed to Android parental control apps, \windowsapps{} operate with more privileges, and use the following techniques:
(1) TLS-interception: a proxy is installed by inserting a self-signed certificate in the trusted root certificate store. This allows the \windowsapps{} to perform content analysis and alter content from HTTPS webpages. 
(2) Application monitoring: user applications are monitored for their usage and duration.
(3) User activity monitoring: some \windowsapps{} take screenshots, record keystrokes, and access the webcam. 

\subhead{\extensions{}}
With appropriate permissions, a parental control extension can use the Chrome API and retrieve the URL contacted by the user, intercept and redirect traffic, read and modify page content and meta-data including cookies.
\subsection{Threat Model}
\label{threatmodel}
We consider the following attacker types with varying capabilities.
(1) On-device attacker: a malicious app with limited permissions on the child/parent device.
(2) Local network attacker: an attacker with direct or remote access to the same local network as the child device. This attacker can eavesdrop, modify, and drop messages from the local network. 
(3) On-path attacker: a man-in-the-middle attacker between the home network and a solution's backend server. 
(4) Remote attacker: any attacker who can connect to a solution's backend server. Attacks requiring physical access to either the child/parent device are excluded from our threat model. \looseness=-1

\subsection{Potential Security and Privacy Issues}\label{ref:sec-issues}
We define the following list of potential security and privacy issues to evaluate \pcsproducts{} (tested using only our own accounts where applicable). 
This list is initially inspired by previous work~\cite{faesslersafer1, reaves2017mo, de2016killed, shasha2018playing}, and then iteratively refined by us.

\begin{enumerate}[noitemsep,topsep=0pt,parsep=0pt,partopsep=0pt,leftmargin=5mm]

\item \emph{Vulnerable client product}: 
A parental control product (including its update mechanism) being vulnerable, allowing sensitive information disclosure (e.g., via on-device side-channels), or even full product compromise (e.g., via arbitrary code execution). 

\item \emph{Vulnerable backend}: The use of remotely exploitable outdated server software, and misconfigured or unauthenticated backend \apiEp{}s (e.g., Google Firebase~\cite{firebase} in \androidapps{}).

\item \emph{Improper access control}:
Failure to properly check whether the requester owns the account before accepting queries at the server-end (e.g., insecure direct object reference). 

\item \emph{Insecure authentication secrets}: 
Plaintext storage or transmission of authentication secrets (e.g., passwords and session IDs).

\item \emph{SSLStrip attack}:
The \pcsproduct{}'s online management interface is vulnerable to SSLStrip attack, possibly due to lack of HSTS enforcement (cf.~\cite{hsts-ndss19, li2017hsts}) .

\item \emph{Weak password policy}: Acceptance of very weak passwords (e.g., with 4 characters or less).

\item \emph{Online password brute-force}: No defense against unlimited login attempts on the online parental login interface.

\item \emph{Uninformed suspicious activities}:
No notifications to parents about potentially dangerous activities (e.g., the use of parental accounts on a new device, or password changes). 

\item \emph{Insecure PII transmission}: PII from the client-end is sent without encryption, allowing an adversary to eavesdrop for PII.

\item \emph{PII exposure to third-parties}: 
Direct PII collection and sharing (from client devices) with third-parties.

\end{enumerate}

\subsection{Selection of Parental Control Solutions}
We chose solutions used in the most popular computing platforms for mobile devices (Android), personal computers (Windows), web browser (Chrome), and selected network products from popular online marketplaces (Amazon).\footnote{As of May 2020, current market shares according one estimate (\url{https://gs.statcounter.com}) are: Android 72.6\%, Windows 77\% and Chrome 63.9\%.} 
We used \quotesd{Parental Control} as a search term on Amazon and Chrome Web Store and selected eight devices and ten extensions. 
For \windowsapps{}, we relied on rankings and reviews provided by specialized media outlets (e.g.,~\cite{techradarreport,pcmagreport,digitaltrendsreport}), and selected eight applications. 
For \androidapps{}, we searched the following terms on Google Play: \quotesd{Parental Control} and \quotesd{Family Tracker}.  
From a total of 462 \app{}s, we selected \selecteddownloaded{} \app{}s with over 10K+ installations, and analyzed them automatically. We also downloaded the companion \app{}s for four \networkdevices{} (\meetcircle{} companion \app{} was already in our dataset as it had 50K+ installs). For six of these \app{}s, the developers made available (via their official websites) alternative APKs with additional features. These APKs were also included in the set of automatically analysed \app{}s, adding up to \totalselected{} \app{}s. 
We installed these \app{}s on an Android phone and removed 15 unresponsive/unrelated \app{}s, making the total of \app{}s analyzed to \totalevaluated{}; 51/\totalevaluated{} are pure \purechildrenapps{}; 24 are pure \parentapps{}; and 78 are used for both parent and child devices, which we termed as \quotesd{\mixedchildparentenapps}. For in-depth analysis, we picked  \numofandroidappsDA{}  popular \androidapps{} representing \numofandroidsol{} parental control solutions.  Each solution may include child app(s), parent app(s), and online parental dashboard.

\section{Methodology}\label{sec:methodology}
\begin{figure*}[htb]
 \centering
 \includegraphics[scale=0.65]{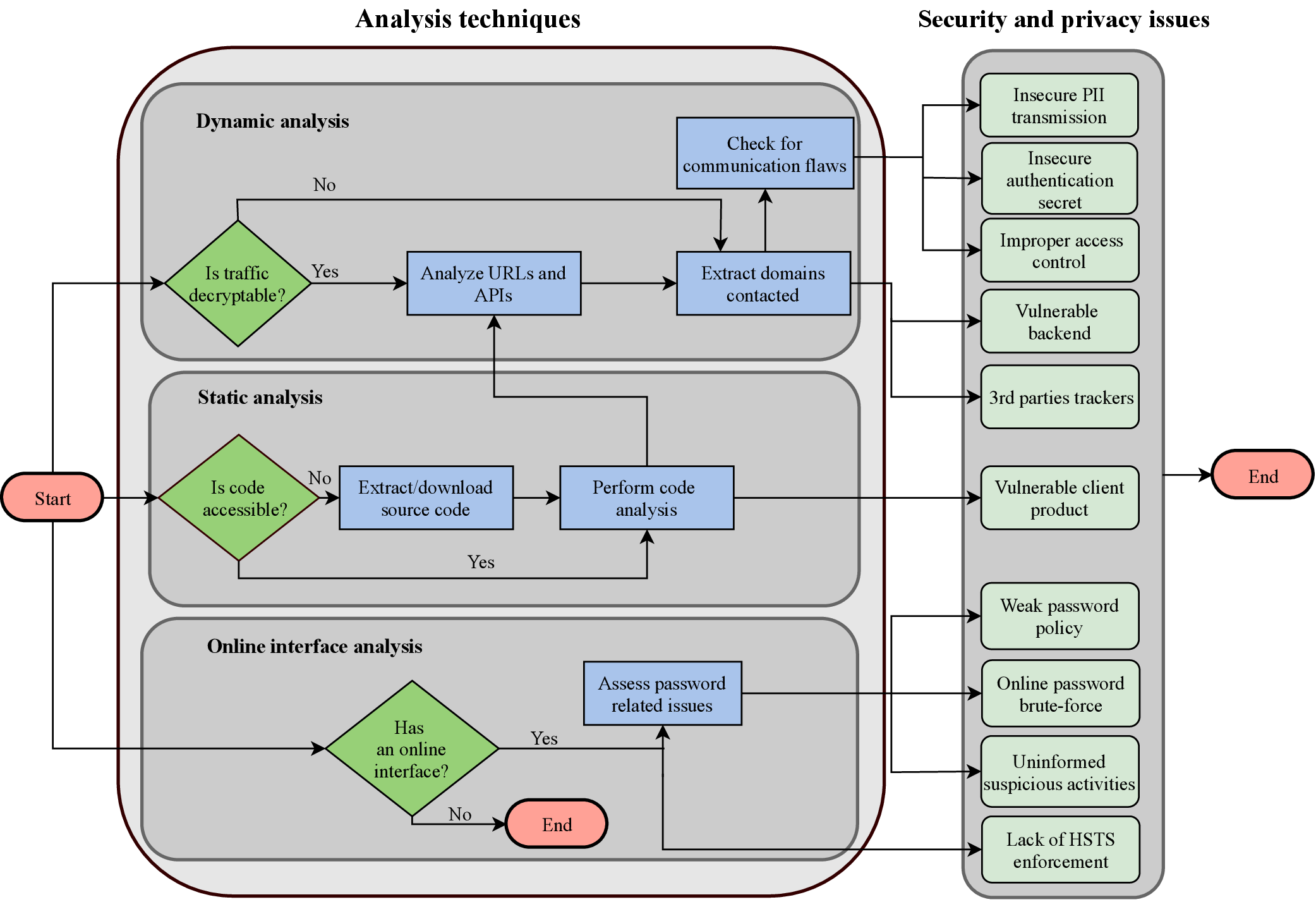}
 
 \caption{Overview of our evaluation framework.}
 \label{fig:methodology}
\end{figure*}

We combine dynamic (primarily traffic and usage) and static (primarily code review/reverse-engineering) analysis to identify security and privacy flaws in \pcsproducts{}; for an overview, see Fig.~\ref{fig:methodology}.
For each product, we first conduct a dynamic analysis and capture the \pcsproduct{} traffic during its usage (as parents/children); if the traffic is in plaintext or decryptable (e.g., via TLS MITM), we also analyze the information sent. 
Second, we statically analyze their binaries (via reverse engineering) and scripts (if available). We pay specific attention to the \apiReq{}s and URLs present in the code to complement the dynamic analysis. After merging the findings, we look into the domains contacted and check the traffic for security flaws (e.g., TLS weaknesses). Third, we test the security and privacy issues described in Sec.~\ref{ref:sec-issues} against the collected API URLs and requests.
Lastly, in case the \pcsproduct{} presents an online interface, we assess the password-related issues and test the SSLStrip attack against the login page.

 \begin{figure*}[htb]
 \centering
 \subfigure[ARP poisoning case.]
 {\includegraphics[scale=0.6,  trim={8.3cm 6.8cm 7cm 10.3cm},clip]{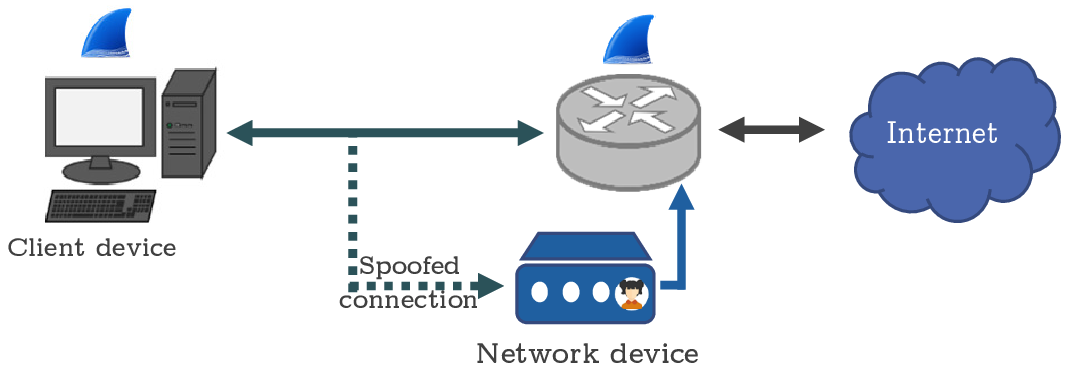}}\quad
 \subfigure[Dedicated Access Point case.]{\includegraphics[scale=0.6,  trim={8.3cm 11.3cm 8cm 6.3cm},clip]{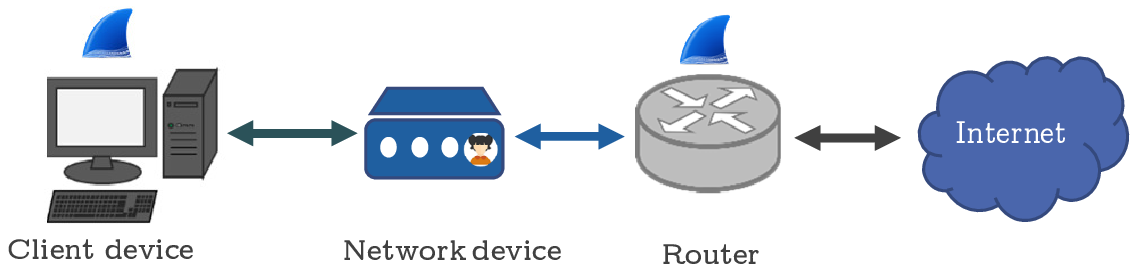}}
 \caption{\Networkdevices{} test environment. Wireshark is installed on both the client device and home router.}
 \label{fig:connecteddevices}
 %\vspace{-10pt} 
 \end{figure*}

\subsection{Dynamic Analysis}
We set up test environments for each solution, emulate user actions for hours to days, collect the traffic from the child, parent, and \networkdevices{}, and then perform relevant analysis (see Sec.~\ref{ref:sec-issues}). 
 
\subsubsection{Usage Emulation and Experimental Setup}
We analyze each solution by manually mimicking regular users' operations with the goal of triggering parental control mechanisms. We test for potential vulnerabilities in these mechanisms (see Sec.~\ref{trafficanalysis}). 
We evaluate the web filtering mechanism by visiting a blocked website (gambling/adult) and a university website. 
We also perform user activities monitored by platform-specific parental control features (see Sec.~\ref{sec:montoring-tech}, and Table~\ref{table:moniteringtechniques} in the Appendix), and evaluate the solution's operations. For example, on Android, we perform basic phone activities (SMS, phone call) and internet activities (Instant messaging, social media, browsing, and accessing blocked content). 

The \networkdevices{} are evaluated in a lab environment by connecting them to an internet-enabled router (like in a domestic network setup) with the OpenWrt firmware~\cite{OpenWrt}. We use test devices with web browsing to emulate a child's device. If the parental control device uses ARP spoofing, the test device is connected directly to the router's wireless access point (AP); see Fig.~\ref{fig:connecteddevices}~(a). Otherwise, the test device is connected to the parental control device's wireless AP; see Fig.~\ref{fig:connecteddevices}~(b). We capture network traffic on both the test device and the router using Wireshark and tcpdump, respectively. 

For \androidapps{}, we maintain two experimental environments to concurrently record and inspect network traffic originating from the child and parent apps. We examine the child apps using a Samsung Galaxy S6 phone running Android 7.0; for the parent apps, we use a Nexus 4 with Android 5.1.1. We run a full Linux distribution with mitmproxy~\cite{mitmproxy} and tcpdump on each experimental environment by installing Linux Deploy~\cite{linuxdeploy}, and configured Android's network settings to proxy all traffic going through the WiFi adapter to the mitmproxy server. This enables us to capture the network traffic directly within the mobile devices. 

We test each \windowsapp{} and \extension{} on a fresh Windows 10 virtual machine with Chrome, tcpdump and mitmproxy installed. We intercept inbound and outbound traffic using mitmproxy on the host, and record packets using tcpdump.

\subsubsection{Traffic Analysis}
\label{trafficanalysis}
After intercepting traffic, we parse and commit the collected tcpdump traffic to an SQLite database and check for the following security and privacy related issues.
 
\subhead{PII and authentication secrets leakage} We examine the collected traffic to check for PII and authentication secrets transmitted in plaintext, or leakage of PII to third-party domains. 
We create a list of possible PII (see Table~\ref{table:list_pii} in the Appendix) that can be leaked via the Request URL, Referer, HTTP Cookie, requests' payload, and LocalStorage. We automatically search for PII items (i.e., case insensitive partial string match) in the collected traffic, and record the leaked information, including the HTTP request URL. We decode the collected network traffic using common  encoding (base64 and URL encoding) and encode possible PII using hashing algorithms (MD5, SHA1, SHA256, and SHA512) to find out obfuscated leaks.

\subhead{Improper access control}
We parse the traffic to find \apiEp{}s with improper access control. First, we try to identify all the APIs that can be potentially exploited (without strong authentication), using Postman (\url{postman.com}) to replay the recorded HTTP request stripped of authentication headers (e.g., cookies and authorization header). Any request successfully replayed is labeled as potentially vulnerable (in a database). Afterwards, we retrieve the parameters used by these APIs (e.g., keys, tokens, or unique IDs), and assess the parameters in terms of their predictability and confidentiality. For instance, we deem a device's access control insecure if its own MAC address is used for \apiEp{}s authentication, as the MAC address can easily be found by an attacker on the local network. \looseness=-1  

\subhead{Identifying trackers}
We use the EasyList~\cite{EasyList}, EasyPrivacy~\cite{EasyPrivacy}, and Fanboy~\cite{Fanboy} to identify known trackers. We also add known trackers  from past work~\cite{vallina2016tracking,razaghpanah2018apps} to our list. 
To identify third-party \domains{} in the \pcsproducts{} traffic, we use the WHOIS~\cite{whois} registration record to compare the \domain{} owner name to the parental control website owner. In cases where the \domain{} information is protected by the WHOIS privacy policy, we visit the \domain{}'s domain to detect any redirect to a parent site; we then lookup the parent site's registration information. If this fails, we manually review the \domain{}'s ``Organization'' in its TLS certificate, if available. Otherwise, we try to identify the \domain{} owner by searching in \url{crunchbase.com}. 

\subsubsection{Backend Assessment}
\label{backendassessment}
Due to ethical/legal concerns, we refrain from using any invasive vulnerability scanning tools to assess backend servers. Instead, we look into the backends' software components as disclosed by web servers or frameworks in their HTTP response headers, such as \quotesd{Server} and \quotesd{X-Powered-By}. We then match these components against the CVE database to detect known vulnerabilities associated with these versions. Additionally, we use the Qualys SSL Test (Qualys 2020: \url{ssllabs.com}) to evaluate the security of the SSL configuration of the \pcsproducts{}' backends.

\subsubsection{Challenges}
During the interception and traffic analysis phase, we encountered several challenges. We summarize them here, including the tools and techniques we use to address them.

\subhead{Network traffic attribution} On \androidapps{}, a key issue is to identify the process that generated the traffic in the absence of the packets' referral metadata. We test how the app behaves when the child uses her device normally (e.g., phone calls, messaging, browsing). These activities produce a large amount of traffic that we need to match to the corresponding processes. 
We use the mitmproxy addon~\cite{mitmproxy} to call \texttt{netstat} to detect the process name for every packet. 
We directly use netstat from the underlying Linux kernel (in our Linux Deploy setup) to capture the process ID and process name as soon as a connection is created, while previous work~\cite{le2015antmonitor,razaghpanah2015haystack} read and parse the system \texttt{proc} directory from the Android Linux kernel by checking the directory periodically. 
This past approach misses connections that are opened and closed before the next time they check the \texttt{proc} directory, while our approach looks into the live connection as soon as a connection is created. We may only miss very short-lived connections that are not detected by netstat. 
To the best of our knowledge, we achieve more reliable traffic-process attribution compared to past work. We leave a full evaluation of the effectiveness of the technique to future work.

\subhead{Traffic interception}
Most \networkdevices{} use TLS for communicating with their backends. This prevented us from inserting a root certificate on these devices, so some of the network traffic generated by them is completely opaque to us. In these cases, we rely on static analysis of the device's firmware. In cases where an \androidapp{} uses certificate pinning to refuse server certificates
signed by any CA other than the pinned certificate in the app, we use SSLUnpinning~\cite{sslunpinning} to attach several hooks in the SSL classes in order to bypass this feature and intercept the communication. 
In cases where the child app installs a VPN on the child device to filter and block websites, we intercept the traffic by deleting the VPN configuration from Android setting on the child device. If the app stops functioning without the VPN, we update the app configuration file whenever possible to disable the setup of the VPN on startup of the app on the child device. 
One \windowsapp{}, \qustodio{} uses its own encrypted certificate store, for which, we extract the associated TLS proxy private key by dumping the process memory.

It is possible that due to our employed measures for traffic analysis and attribution (e.g., rooted device, disabled VPN), some parental control solutions may have functioned differently, which is difficult to verify due to the use of heavily obfuscated code. Hence, our findings may be the lower-end of the actual privacy exposure.

\subsection{Static Analysis} 
Our static analysis aims to complement the dynamic analysis whenever we could not decrypt the network traffic (e.g., in case of network devices using TLS). We use static analysis to identify PII leakage, contacted domains, weak security measures (e.g., bad input sanitization), or potential flaws in implemented mechanisms.

\subhead{\Networkdevices{}} We analyze the \networkdevice{} firmware whenever possible. We either attempt to extract the firmware directly from the device (via JTAG, UART, or ICSP interfaces), or download the device firmware from the vendor's website. We found 3/8 \networkdevices{} with an accessible serial UART port (KoalaSafe, Blocksi, and Fingbox) that we used to extract the firmware from the devices.
Another device (Circle) made its firmware available online. Among the remaining devices (without access to their firmware), we scan for the presence of open remote admin services (e.g., SSH), which are often closed or key-protected. 
To identify vulnerable services, we scan the \networkdevices{} with several tools (OpenVas, Nmap, Nikto~\cite{Nikto} and Routersploit~\cite{Routersploit}), and match the identified software versions against public vulnerability databases.

\subhead{\extensions{}}
We manually analyze the source code of the \extensions{}, which mainly consists of scripts, separated into content scripts and background scripts. 
As most \extensions{}' codebase is relatively small, and do not involve serious obfuscation, we can investigate their operations and detect security and privacy issues (e.g., PII leakage, common JavaScript vulnerabilities).

\subhead{\androidapps{}}
We perform an automated analysis on all \totalevaluated{} \androidapps{} using Firebase Scanner~\cite{firebasescanner} 
to detect security misconfigurations in Firebase.\footnote{Google Firebase (\url{https://firebase.google.com/}) provides support for backend infrastructure management for \androidapps{}.} 
 We also use LibScout~\cite{backes2016reliable} to identify third-party libraries embedded in these \app{}s. 
Since LibScout does not distinguish which libraries are used for tracking purposes, we use Exodus-Privacy~\cite{exodus} to classify tracking \SDKs{}. We use MOBSF~\cite{MOBSF} to extract the list of third-party tracking \SDKs{} from all \totalevaluated{} \app{}s based on Exodus-Privacy's tracker list.

\subsection{Online Interface Analysis}
The online user interface is the primary communication channel between parents and \pcsproducts{}. 
It displays most of the data collected by the solutions, and may remotely enable more intrusive features. Compromising the parent account can be very damaging, and thus we evaluate the security of this interface.

 \subhead{SSLStrip attack}
 To check for SSLStrip attacks, we first set up a WiFi AP with mitmproxy, SSLStrip2~\cite{sslstrip2} and Wireshark installed. Then, we connect the \pcsproduct{} to our WiFi access point. Wireshark is utilized to record network traffic while mimicking common use case scenarios with the goal of triggering all parental control monitoring and control UI and \apiReq{}s looking for signs of successfully SSL Stripping attack on the traffic. We confirm the effectiveness of the attack by comparing the result to the corresponding traffic in a regular testing environment (i.e., without SSLStrip).

\subhead{Weak password policy}
During the \pcsproduct{}'s account creation, we evaluate its password policy. 
We adopt a fairly conservative stance and only labelled as weak the password policy accepting password with 4 characters or less.

\subhead{Online password brute-force}
We use Burp Suite~\cite{burpsuit} to perform password brute-force attacks on our own online accounts. To keep the load on the server minimal, we test for the presence of defensive mechanisms by 50 attempts on our account from a single computer.

\subhead{Uninformed suspicious activities}
To determine whether the solution presents measures to report suspicious activities, we test two scenarios in which the user should be notified: modification of the user's password, 
and connection to the account from a new/unknown device. 
We deem a \pcsproduct{} that does not alert (e.g., via email) in either case to be vulnerable.

\begin{table*}[ht]
\centering
\caption{Overall results for security flaws in \pcsproducts{} labelled following the threat model in Sec.~\ref{threatmodel}. \Circle~: On-device attacker; {\protect\includegraphics[width=6.3pt]{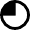}}~: Local network attacker; \LEFTcircle ~: On-path attacker; \CIRCLE~: Remote attacker; \formatSpecial{-}: not applicable; \formatSpecial{blank}: no flaw found. In case the vulnerability can be exploited by 2 types of attackers, we display the fullest circle applicable.
}
\label{table:overallvuln}
\begin{tabular}{@{}l}
\includegraphics[]{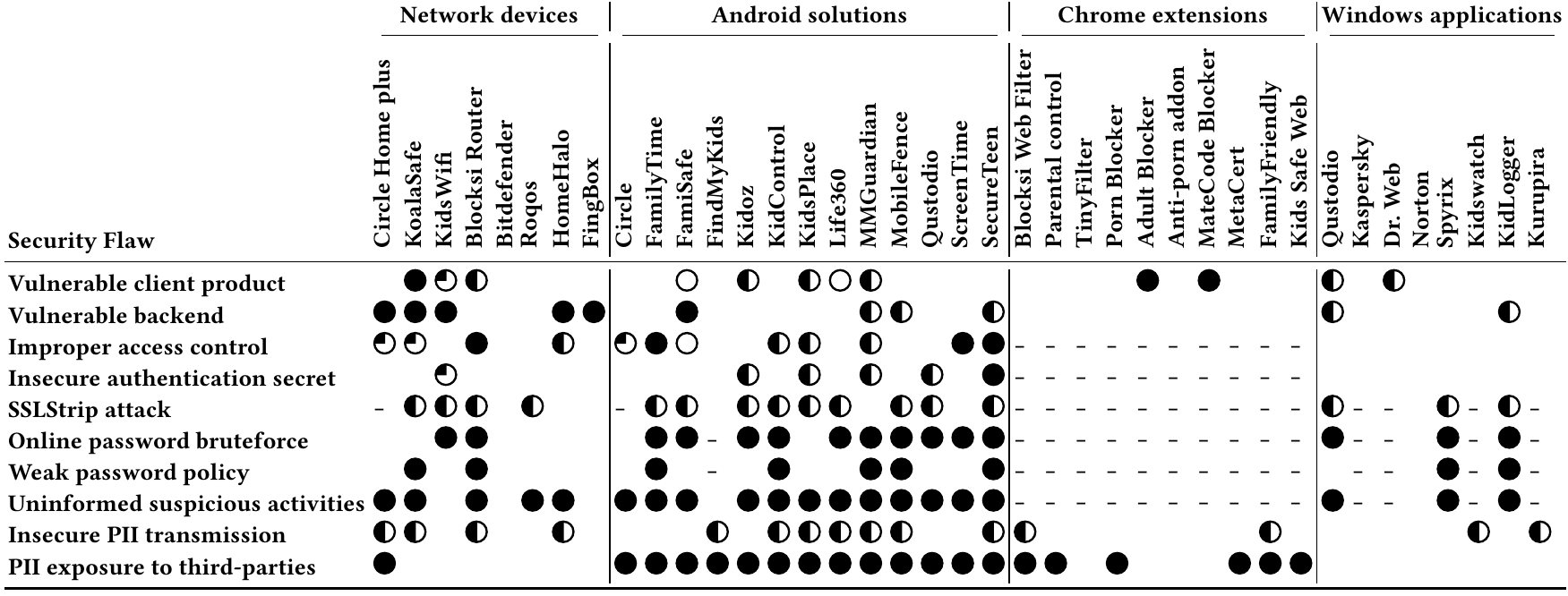}
\end{tabular}
\end{table*}

\section{Results}

Following the methodology in Sec.~\ref{sec:methodology}, we analyzed the \pcsproducts{} between \testPeriod{}, which include: 8 \networkdevices{}, \numofandroidappsDA{} \androidapps{} representing \numofandroidsol{} \androidsols{}, 10 \extensions{} and 8 \windowsapps{}. We also performed an automated analysis of \totalevaluated{} parental control \androidapps{} to detect vulnerable backend databases and check for tracking SDKs.
In this section, we report our findings on the tested security and privacy issues (as outlined in Sec.~\ref{ref:sec-issues}); for an overview, see Table~\ref{table:overallvuln}.

\subsection{Vulnerable Client Product}
\label{devVulns}

\subhead{\Networkdevices{}}
The importance of securing the update mechanism has been known for years, cf.~\cite{sw-update-2006}. Surprisingly,
the \blocksi{} firmware update happens fully through HTTP. An integrity check is done on the downloaded binary image, using an unkeyed SHA256 hash, again retrieved using HTTP, and thus rendering it useless. Therefore, an on-path attacker can trivially alter the update file and inject their own malicious firmware into the device. We confirmed this vulnerability to be exploitable. We also found another vulnerability 
that enables executing a command as root on the Blocksi device via command injection (i.e., unsanitized user input is passed directly to a system shell for execution). 
We confirmed this vulnerability to be exploitable by sending a \texttt{router\_setGeneralSettings} request to the \blocksi{} \apiEp{}, and injecting a command in the timezone field in the request parameters. The settings change triggers a WebSocket Secure (WSS) message to the \blocksi{} device. The device then reads the new configuration from the \apiEp{} and updates its local configuration.\footnote{The timezone value is passed as \texttt{tz} to [``echo'' + tz + ``> /etc/TZ'']. Thus, if \texttt{tz} is ``\$(ls)'', the \texttt{ls} command would be executed and its output written to /etc/TZ.}

We also found that KoalaSafe runs Dropbear v2014.63 SSH server/client (released on Feb.\ 19, 2014), associated with four known remote code execution vulnerabilities. Under certain conditions, the KoalaSafe device opens a reverse SSH tunnel through its backend server, exposing the vulnerable SSH Dropbear server to an attacker outside the local network. By calling a KoalaSafe \apiEp{},\footnote{\url{https://api.koalasafe.com/api/router/[MAC address]/et}} an external attacker can detect when a reverse SSH tunnel is open using only the victim device's MAC address. If the tunnel is open, the \apiEp{} responds with the tunnel's port number, 0 otherwise. 
For large-scale exploitation, an attacker can query the aforementioned \apiEp{} to enumerate all KoalaSafe devices with the reverse tunnel open.
This enumeration is feasible as KoalaSafe uses the GuangLia network interface card (NIC), and MAC addresses assigned to GuangLia NICs~\cite{GuangLia} are limited to only $2^{20}$ values.

\subhead{\androidapps{}}
We found 3/\numofandroidsol{} \androidsols{} (\famisafe, \kidsplace{} and \life{}) do not encrypt stored user data on shared external storage that can be accessed by any other apps with the permission to access the SD card. Examples of the sensitive information include: the parent's email and PIN code, phone numbers, the child’s geolocation data, messages and social media chats, visited websites, and even authentication tokens---which enabled us to read private information from the child account remotely. 

We also found that \kidoz{}, \kidsplace{}, and \mmguadian{} use custom browsers to restrict and filter web content. The three browsers fail to enforce HSTS, and lack persistent visual indication if the website is served on HTTP. \kidsplace{} safe browser keeps the address bar that shows visited URL to help with visual identification. However, \mmguadian{} shows the URL in the address bar until the page is fully loaded and then the URL is replaced with the webpage title. Following our disclosure, \mmguadian{} removed their custom browser.\looseness=-1

\subhead{\windowsapps{} and \extensions{}}
Other than Kidswatch, all tested \windowsapps{} relied on TLS proxies to operate. Some of these proxies do not properly perform certificate validation. For example, Qustodio and Dr.\ Web accepted intermediate certificates signed with SHA1, despite the enhanced collision attack on SHA1~\cite{leurent2019collisions}. Dr.\ Web also accepted Diffie-Hellman 1024 (considered weak~\cite{weakdh15}, and deprecated in Safari and Chrome since 2016~\cite{DHEChrome}). 
In addition, none of the proxies rejected revoked certificates.
We also found that upon uninstallation of these applications, the root certificates associated with the proxies remained in the Windows trusted root certificate store, with four of them having a validity duration over one year.

Two Chrome extensions (Adult Blocker and MateCode Blocker) download and run a third-party tracking script at run time. The domains hosting the scripts are not apparently related to the extension providers (or libraries from well-known companies). Note that runtime loaded scripts bypass the static control of Chrome for extension security, which has been exploited in the wild by tricking developers into adding malicious scripts masquerading as tracking scripts~\cite{extension-attack}.\looseness=-1

\subsection{Vulnerable Backend}
\label{vulnerablebackend}

\subhead{\Networkdevices{}}
Examples of vulnerable software components from our analysis of backend server \apiEp{}s include: Apache 2.4.34 with 11 CVEs in KoalaSafe; PHP 7.0.27 with  26 CVEs in KidsWifi; Nginx versions with the same 3 CVEs in KidsWifi, Circle, HomeHalo and Fingbox. 
The \blocksi{}'s \apiEp{} only indicates that it runs on \href{https://openresty.org/en/}{OpenResty} and Google Frontend (no version info). 

\subhead{\androidapps{}}
Since 115/\totalevaluated{} \androidapps{}   use Google Firebase as a backend service, we analyzed their Firebase configuration for security issues by performing an automated analysis using Firebase Scanner~\cite{firebasescanner}. 
Critical misconfigurations can allow attackers to retrieve all the unprotected data stored on the cloud server. We followed a similar approach to Appthority's work~\cite{Firebasevulnerability} on scanning \app{}s for Firebase misconfigurations. We found 8/\totalevaluated{} \androidapps{} with insecure Firebase configurations. We then evaluated the type of sensitive data exposed by each app to determine the impact of the data being leaked. 
For ethical reasons and to protect other customers privacy, we created a parental account on the eight \app{}s. Then, we updated the Firebase scanner to automatically search for our test data in the its response and record the leaked information from our own account. We found three \app{}s exposing personal information: 1) \famisafe{} with 500K+ installs exposes the parent email; 2) \href{https://play.google.com/store/apps/details?id=com.careapps.locate}{Locate} with 10K+ installs exposes the child name, phone number, and email; and 
3) \href{https://play.google.com/store/apps/details?id=com.ailemonline.mobile}{My Family Online} with 10K+ installs exposes the child name, child and parent phone numbers, parent email, and apps installed on child phone. FamiSafe fixed the Firebase  security issue following our disclosure.
Additionally, we found that \mmguadian{}, \mobilefence{}, and \secureteen{} servers support RC4, and \secureteen{} backend is vulnerable to the POODLE attack.\looseness=-1

\subhead{\windowsapps{}}
We found that some \windowsapps{}' servers also do not use ideal 
TLS configurations. For instance, Qustodio's server has an intermediate certificate signed with SHA1 in its chain of trust. Qustodio and KidLogger servers support the RSA key exchange protocol which lacks forward secrecy.

\subsection{Improper Access Control}
\label{sectionInfoDisclosure}

\subhead{\Networkdevices{}}
The KoalaSafe API login endpoint requires three parameters that are available to anyone on the local network: a device-generated authentication token, the device’s date and time, and the device’s MAC address for successful authentication. These parameters can be obtained by visiting endpoints hosted by the KoalaSafe device.\footnote{Authentication token and device time are available at \url{https://device.koalasafe.com/auth.lua}, and the MAC address at \url{https://device.koalasafe.com/status.lua}} Thus, a local network attacker can easily collect the information needed for authentication and use the \apiEp{} to access sensitive information such as the profile name, email address, and browsing history. 

For \blocksi{}'s login \apiEp{}, the device's serial number (SN) and the registered user's email are required to authenticate the device to the server. However, a remote attacker needs to know only one of these parameters to authenticate. This is because a remote attacker can retrieve a user's email using their device SN or vice-versa.\footnote{For SN to email, use
\url{https://service.block.si/config\_router\_v2/router\_checkRouters/null/[SN]},
and for email to SN, \url{https://service.block.si/config\_router\_v2/router\_checkRouters/[email]}.} By sending both parameters to the \apiEp{} in a POST message, any remote attacker can authenticate to the server, and access sensitive information about the home network, e.g., the WiFi password, and MAC addresses of connected devices. 

The HomeHalo device uses only the device's SN and an HTTP header called \texttt{secretToken} to authenticate to its \apiEp{}. In our case, the secretToken had a fixed value of 100500. 
An on-path attacker can intercept and modify these messages, and gain access to admin controls, e.g., reading or changing the wireless SSID, password, or even the device's root password. Other privacy sensitive information is also exposed, including: the devices connected to HomeHalo's network and the parental control profile setup.\looseness=-1

The Circle Home Plus creates a profile for each child and stores it locally on the device, including the child age groups, usage history and statistics, child photo, and username (i.e., some parents may use child name).
We identify two \apiEp{}s used to transmit child information in plaintext over the local network. 
The first \apiEp{}\footnote{ \url{http://10.123.234.1/api/USERINFO?host=ios\&nocache=1572292313630 HTTP/1.1}} sends child account usage history and statistics, and \texttt{profileID}. It insecurely relies on the requester’s MAC address to identify the child device and communicate sensitive information. This \apiEp{} is called whenever a child device attempts to access a restricted domain. The second \apiEp{}\footnote{\url{http://10.123.234.1/api/USERPHOTO?profileID= [profileID]}.} fetches the profile photo corresponding to the received profile ID.

\subhead{\androidapps{}}
We  found 8/\numofandroidsol{} \androidsols{} lack authentication for accessing PII.
Prominent examples include the following.
In \familytime{}, a six-digit parameter \texttt{childID} is generated through a sequential counter incremented by one per user. An attacker can retrieve the child name, gender, date of birth, email address, and child phone number through an \apiReq{} that requires only the childID value. Hence, an attacker can remotely exploit this vulnerability at a large scale, simply by trying all 6-digit values.\footnote{By using e.g., a cURL commad (the last parameter is childID): 
\$ curl -v \url{https://mesh.familytime.io/v2/child/Android/profile/456***}.}
In \famisafe{}, an attacker can retrieve all the child social media messages and YouTube activities labeled as suspicious through an \apiReq{} that requires the following parameters: \texttt{deviceid}, \texttt{memberid}, \texttt{client\_sign}, and \texttt{access token}. However, any app installed on the child device can access these parameters from the \famisafe{} log file on the shared external storage.\footnote{By using e.g., a cURL command:
\$ curl -v \url{https://u.famisafe.com/load-page /index?page=suspicious-text/detail\&access \_from=1\&device\_id=165***\&member\_id=1045***\&client\_sign={fffff***-be**-19ec-0000-000075b3****}\&access\_token=dtwMtFarI********\&lang=en}.}

\subsection{Insecure Authentication Secret}
\label{insecureauthensecret}

\subhead{\Networkdevices{}}
During the setup procedure of \kidswifi{}, the device creates an open wireless AP  with SSID ``set up kidswifi'', making it temporarily vulnerable to eavesdropping. The parent has to use this AP's captive portal to configure the \kidswifi{} device to connect to the home network. 
Consequently, as this AP is open and the client-device communication happens through HTTP, the home router's WAN and \kidswifi{}'s LAN credentials become available to local attackers. We deem this a minor risk as the vulnerability is only present for a limited duration (during device setup), and the attacker must be within close proximity.

\subhead{\androidapps{}}
In \secureteen{}, we found an \apiEp{} that can be used to authenticate the user to the parental control account. This \apiEp{} enables any adversary to remotely compromise any parental account by knowing only the parent's email. When the \apiReq{} is invoked by the browser, the adversary is logged in to the parental dashboard and obtains full access to the parent account, including the ability to monitor and control the child device.\footnote{An example call to the API is as follows: \url{https://cp.secureteen.com/auth.php?\&productName=secureteen\&resellerId=careteen\&page=menu\&loginFromApp=Yes\&j\_username=parentemail**@gmail.com\&gType=monitoring}.}

\kidoz{} exposes the user email and password in HTTP when the \quotesd{Parental Login} link is clicked from the \href{https://kidoz.net/}{https://kidoz.net} home page. 
\kidsplace{} and \qustodio{} leak session authentication cookies via HTTP, with validity periods of one year and two hours 
respectively. Even with the 2-hour cookie in \qustodio{}, the attacker can easily access sensitive information about the child including the child's current location, and history of movements. The attacker can also access remote control functions on the child phone, such as block all incoming/outgoing calls. In the case of \kidsplace{}, the attacker can access a wide spectrum of remote control functions to the child phone such as: disable the Internet, silently install a malicious app on the child device, or upload harmful content to the child mobile. The attacker can also lock the child phone making her unable to contact the parent or perform an emergency call.

\subsection{SSLStrip and Online Account Issues}
We found that nine \androidsols{}, four \networkdevices{}  and three \windowsapps{} transmitted 
the parent account credentials via HTTP under an SSLStrip attack.
This allows an adversary to compromise the parent account for a long time, particularly if the app does not send any notification to the parent when the account is accessed from a new device.
More seriously, in \kidoz{}, we could see the parent's credit card account number and email in HTTP when using their BlueSnap online payment solution~\cite{bluesnap}, while connected to our WiFi access point. This was possible because the online payment server is not configured to use HSTS. 
In \qustodio{}, we could extract the child Facebook credentials provided by the parent during the configuration of the monitoring component. Following our disclosure, \familytime{} enabled HSTS on their server.

In terms of defense against online password guessing, we found that two \networkdevices{} and 10 \androidsols{} leave their online login interfaces open to password brute-force attacks. Also, two \networkdevices{}, five \androidsols{}, and three \windowsapps{} enforced a \emph{weak} password policy (i.e., shorter than four characters). We also observed that five \networkdevices{}, 12 \androidsols{ } and four \windowsapps{} do not report suspicious activities on the parent's account such as password changes and accesses from unrecognized devices. These activities are possible indicators of account compromise and should be reported to the user.\looseness=-1

\subsection{Insecure PII transmission}

\subhead{\Networkdevices{}} We found that the KoalaSafe and \blocksi{} \networkdevices{} append the child device's MAC address, firmware version number, and serial number into outgoing DNS requests. This can allow on-path attackers to track the child's web activities~\cite{mac-leakage}.
The HomeHalo device suffers from a similar problem: whenever a domain is requested by a user device inside its network, HomeHalo sends an HTTP request, including the child device's MAC address, to its backend server to identify the requested domain's category.

\subhead{\androidapps{}}
Several \androidsols{} send cleartext PII, see Table~\ref{table:leakviahttp} in the Appendix. Examples include:
\findmykids{} (the child's surrounding sounds and photo);
\KidControl{} (the parent's name and email, geolocation, and SOS requests); and
\mmguadian{} (the parent's email and phone number, and child's geolocation). 
\mmguadian{}  transmits the child visited URL (Base64 encoded) to a third-party domain classifier \url{Komodia.com}~\cite{komodia}) via HTTP. When we contacted MMGuardian, they informed us that they are working with Komodia on a resolution. Other products using Komodia are also apparently affected by this.

\subhead{\windowsapp{} and \extensions{}}
During the installation phase of Kurupira, the user has to set up an SMTP server with the assistance of the application to receive activity reports. However, in case the user uses an SMTP server with an unencrypted protocol, Kurupira does not warn about transmitting child activity report in plaintext. Kidswatch sends child activity reports over HTTP.
We also found that three extensions (Blocksi Web Filter, FamilyFriendly Parental Control, Porn Blocker) send the domain contacted by the user to the extension’s server using HTTP to check whether or not the website should be blocked.

\subsection{Third-party SDKs and Trackers}
\label{Sec:3rdparties}
Some legislations (e.g., US COPPA and EU GDPR) regulate the use of third-party trackers in the services targeting children (e.g., under 13 years of age). 
We thus evaluate potential use of third-party tracking \SDKs{} in the \pcsproducts{}. 
We found notable use of third-party \SDKs{} in \pcsproducts{}, except in Windows. For \networkdevices{}, we identified the use of third-party \SDKs{} in the companion \app{}s but not in the firmware.

\subhead{Trackers}
In Android, we found use of trackers in most \app{}s via static analysis, including: the \purechildrenapps{} (targeted for children's devices only, 44/51 \app{}s with tracking SDKs), \mixedchildparentenapps{} (the same app is used by both parents and children, 73/78 \app{}s), and \parentapps{} (targeted for parents' devices only, 22/24 \app{}s); see Table~\ref{table:countoftrackingsdks} in the Appendix. Over 25\% of \purechildrenapps{} utilize advertising networks (e.g., Google Ad and Doubleclick \SDKs; see Fig.~\ref{fig:Percstaticanalysis} in the Appendix) which could potentially violate US COPPA.  For \networkdevices{}, our static analysis for five companion \app{}s reveals the use of tracking \SDKs{} (2--12 unique trackers) in all those \app{}s except for KoalaSafe. For \extensions{}, we found that half of the \extensions{} send behavioral information (e.g. web browser usage) to Google Analytics.

We also identify tracking third-party \domains{} from network traffic generated during our dynamic analysis from child device. 
Except \secureteen{}, 12/\numofandroidsol{}  \androidsols{} use tracking \SDKs{} (1--16 unique trackers; see Fig.~\ref{fig:Percdynamicanalysis} in the Appendix). Our traffic analysis confirms violations of COPPA---over 30\% of \androidsols{} utilize \url{doubleclick.net} without passing the proper COPPA compliant parameter from child device.\footnote{ The use of \texttt{tfcd=1} marks an ad request as child-directed; see \url{https://support.google.com/admanager/answer/3671211?hl=en}.} 
We also found that one of the \networkdevices{}' companion app, Circle, includes a third-party analytical SDK from Kochava. Every time the app is launched, or it returns to the foreground, the following information is shared with Kochava: Device ID (enables tracking across apps), device data (enables device fingerprinting for persistent tracking). Kochava provides an opt-out option (\texttt{app\_limit\_tracking=true}) that can be used to comply with COPPA. However, the \meetcircle{}  transmits this flag as \texttt{false} from the child device.\footnote{Note that Disney, a former partner of Circle, is the target of a class action lawsuit for using a similar SDK in children's apps; see \url{https://unicourt.com/case/pc-db1-rushing-et-al-v-the-walt-disney-company-et-al-494632}.}

For \androidsols{} that have a safe custom browser, such as~\kidoz{},~\mmguadian{}, and~\kidsplace{}, we found that all these browsers allow visited websites to store persistent tracking HTTP cookies (or Local Storage) on the child device. These cookies are not erased when the browser \app{} is closed. 

\subhead{Restricted \SDKs{} from past work}
We also study the SDKs identified in past studies~\cite{reyes2018won, feal2019angel} that are restricted by their developers (e.g., fully prohibited, or use with particular parameters) for use in children's apps (as stated in their policies as of June 2020).  
We 
evaluated the privacy policies for the seven prohibited SDKs detected, and concluded that four companies, Crashlytics, Amplitude, Braze (formerly Appboy), and Appnext, still prohibit the use of their \SDKs{} in children's apps; two others (Tapjoy and Branch) now require developers to set the appropriate parameters; and the last one (Supersonic/ironSource) removed any restriction.

From static analysis, we found several prohibited SDKs being used: 25/44 \purechildrenapps{} and 8/73 \mixedchildparentenapps{} use Google CrashLytics; unGlueKids \purechildrenapp{} uses Branch \SDK{} (without the \texttt{do-no-track} mode); and Limitly uses Appnext \SDK{}. 
Aside from Google CrashLytics, we also observed Branch (7 \app{}s), Amplitude (6 \app{}s), Braze (4 \app{}s), and Tapjoy (1 \app{}) \SDKs{} in the \mixedchildparentenapps{}. 

Through analysing traffic generated from child device, we confirm that five \androidsols{} use prohibited \SDKs{}. 
Also for \life{}, we note that Branch \SDK{} \quotesd{do-not-track} mode was  disabled since the network traffic from child device contains Android ID,  Android Advertising ID (AAID), and local private IP. 
Additionally, three \androidsols{} \findmykids{}, \kidsplace{}, and \meetcircle{} contact Crashlytics prohibited \SDK{} server (\url{reports.crashlytics.com}), and \qustodio{} communicates with the Braze prohibited \SDK{}.

\subhead{PII exposure to third-parties}
We found that all \androidsols{} share personal and unique device information with third-party domains (see Table~\ref{table:sharethirdparty} in the Appendix). Prominent examples include:  
\screentime{} shares the child Android ID with Facebook. 
Four extensions send the requested domains to their server to check whether the website should be blocked, which can also be locally performed similar to Google Safe Browsing. More concerning 2/10 extensions send the complete URL, possibly leaking personal information not required for blocking. 
Another extension, Parental Control, overrides Chrome setting and replaces the default search URL by its server domain, which automatically redirects to Google Safe Search, but exposes the search terms to the extension's server. 
We also found that another \extension{}, Porn Blocker,  redirects the user to \url{https://www.purplestats.com/page/blocked/} when visiting a blocked website, and leaks the full URL of the previous webpage through the referer header.

\subhead{COPPA Safe Harbor providers}
We check the behavior of (3/\numofandroidsol{}) (\kidoz{}, \familytime{}, \findmykids{}) \androidsols{} certified by the US FTC's COPPA Safe Harbor program~\cite{safeharborprogram} (by kidSAFE~\cite{familytimesafeseal}; we also checked other programs under Safe Harbor, and the \pcsproducts{} websites/descriptions). 
Our traffic analysis collected from the child device reveals that \findmykids{} use three trackers and leak Android Advertising ID to at least two trackers \url{graph.facebook.com} and \url{adjust.com}. \findmykids{} includes two flags when calling Facebook to enable application tracking and advertiser tracking (both were enabled)~\cite{FacebookAPI}.  
\findmykids{} also shares child Android ID with Yandex Metrica (\url{appmetrica.yandex.net}). Yandex Metrica provides an option  (\texttt{limit\_ad\_tracking}) that can be used to restrict tracking. However, the \findmykids{}  transmits this flag as \texttt{false} from the child device~\cite{Yandex}. 
We also found that \familytime{} sends the child's name, email address and phone \# (hashed in SHA256) to \url{facebook.com}.  
\kidoz{} uses eight trackers and leaks the Android Advertising ID to the third-party domain \url{googleapis.com} through the referer header.

\section{Potential Practical Attacks}
In this section, we summarize the impact of exploiting some of the discovered vulnerabilities in the analyzed {\pcsproducts}.

\subhead{Device compromise}
Device compromise presents serious security and privacy risks, especially if a vulnerability can be exploited remotely. We found multiple vulnerabilities in the \blocksi{} \networkdevice{} that can compromise the device itself.
These include an exploitable command injection vulnerability and a vulnerability in protecting the device's serial number, which is used in authentication.
A remote attacker can use these vulnerabilities to take control over the \blocksi{} device by simply knowing the parent's email address (see Sec.~\ref{devVulns} and Sec.~\ref{sectionInfoDisclosure}). In particular, using the serial number and email, an attacker can exploit the command injection vulnerability 
and spawn a reverse TCP shell on the device. At this stage, the attacker gains full control of the device, and can read/modify unencrypted network traffic, disrupt the router's operation (cf.\ DHCP starvation~\cite{tripathi2015exploiting}), or use it in a botnet (cf.~Mirai~\cite{mirai}).

\subhead{Account takeover}
Parental accounts can be compromised in multiple ways. First, none 
of the \pcsproducts{}' web interface except Norton enforced HSTS, and most were found vulnerable to SSLStrip attacks. Therefore, an on-path attacker can possibly gain access to the parent account using SSLStrip, unless parents carefully check the HTTPS status. Second, login pages that allow unlimited number of password trials could allow password guessing (especially for weak passwords). Note that most \pcsproducts{}' password policies are apparently weak (cf.\ NIST~\cite{grassi2017digital}); some products accept passwords as short as one character. Third, products with broken authentication allow access to parental accounts without credentials. For example, \secureteen{} provides an \apiEp{} (see Sec.~\ref{insecureauthensecret}) to access the parental account, by knowing only the parent email address. 
If logged-in, the attacker has access to a large amount of PII, 
social media/SMS messages, phone history, child location---even enabling possibilities of  physical world attacks. 

\subhead{Data leakage from backends}
Failure to protect the parental control backend databases exposes sensitive child/parent data at a large scale.
Firebase misconfigurations exposed data that belongs to 500K+ children and parents from three \app{}s.  
Such leakage may lead to potential exploitation of children both online and offline.

\subhead{PII on the network}
COPPA mandates reasonable security procedures for protecting children's information~\cite{coppa}. However, we found several \pcsproducts{} transmit PII insecurely. 
For example, \findmykids{} leaks surrounding voice, and the child's picture. This could put a child in physical danger since the attacker can learn intimate details from the child's voice records and her surrounding, and also recognize the child from her photo.
\KidControl{} allows the child to send SOS messages when she is in a dangerous situation. However, an attacker can identify and drop the SOS message at will as it is sent via HTTP.  
Moreover, KoalaSafe and \blocksi{} \networkdevices{} append the child's device MAC address to outgoing DNS requests, enabling persistent tracking. 

\section{Conclusion}
Parental control solutions are used by parents to help them protect their children from online risks. Nevertheless, some of these solutions have made news in the recent years for the wrong reasons. Our cross-platform comprehensive analysis of popular solutions  
shows systematic problems in the design and deployment of 
\emph{all} the analyzed solutions (except Bitdefender, TinyFilter, Anti-porn addon, Kaspersky, and Norton) from a security and privacy point of view. 
Indeed several of these solutions can undermine children's online and real-world safety. As these solutions are viewed as an essential instrument to provide children a safer online experience by many parents, these solutions should be subjected to more rigorous and systematic evaluation, and more stringent regulations.

\begin{acks}
     This work was partly supported by a grant from the Office of the Privacy Commissioner of Canada (OPC) Contributions Program. We thank the anonymous ACSAC 2020 reviewers for their insightful suggestions and comments.
\end{acks} 

\bibliographystyle{ACM-Reference-Format}
\bibliography{references}

\section{Appendix} 
\label{firstappendix}
In this appendix, we first provide some recommendations for parental control solution providers. Then, we present the corpus of \pcsproducts{} that we evaluated. Then, we provide a summary of the  techniques adopted by the analyzed \androidsols{} to monitor child activities. Finally, we report our observations of tracking and PII sharing done by third-party SDKs and libraries embedded in these \pcsproducts{}.

\subsection{Recommendations}

In what follows, we list our recommendations for parental control solution providers. 

\subhead{Addressing vulnerabilities}
Because of the sensitivity of the information manipulated by the \pcsproducts{}, companies should conduct regular security audits; the issues we listed in Sec.~\ref{ref:sec-issues} can serve as a starting point. 
Moreover, they should have a process to address vulnerabilities such as responsible disclosure and bug bounty programs. Currently, none except Kaspersky and Bitdefender participates in such programs.

\subhead{Enforcing best practices}
Parental control companies should rely on publicly available guidelines and best practices, including proper \apiEp{} authentication and web security standards~\cite{owasp, owaspAPI}. We also strongly encourage companies to adopt a strong password policy in their products, because the use of default, weak and stolen credentials has been exploited in many known data breaches~\cite{verizon2019dbir}.
In the case of network devices, manufacturers should employ a secure firmware update architecture (see e.g., IETF~\cite{ietf-suit-architecture-08}). Adopting known best practices is critical due to the especially vulnerable user base of these products.

\subhead{Monitoring account activities}
\Pcsproducts{} should report suspicious activities on the parent's account such as password changes and accesses from unrecognized devices. These activities could indicate account compromise. 

\subhead{Limiting data collection}
\Pcsproducts{} should limit the collection, storage, and transmission of the children's data to what is strictly necessary. For instance, the solution should not store PII not required for the solution's functionality. 
The \pcsproducts{} should also allow the parent to selectively opt-out of the data collection in certain features. 

\subhead{Securing communication}
Transmission of PII should happen exclusively over secure communication channels. 
The solution should utilize MITM mitigation techniques such as host white-listing, certificate pinning, and HSTS~\cite{hsts}.

\subhead{Limiting third-parties and SDKs}
\Pcsproducts{} should limit the usage of trackers and tracking SDKs in \app{}s intended for children. For the SDKs that allow special parameter for children's apps, those parameters must be used appropriately. 

\subsection{Parental Control Solutions Corpus} 
Tables \ref{table:listdevices}, \ref{table:listapp_tab},  \ref{table:listofWindowsapps} and \ref{table:listext_tab} provide some information about the corpus of the parental control solutions analyzed throughout our study.
\begin{table}[H]
\caption{List of parental control devices and their firmware versions.}%  
\begin{tabular}{ll}
\textbf{Device}       & \textbf{Version}      \\\hline
Circle Home Plus      & 3.10.0.2     \\\hline
KoalaSafe    & 1.26825      \\\hline
KidsWifi     & 1.165        \\\hline
Blocksi Router
& 2.4          \\\hline
Bitdefender  & 2.1.66.4     \\\hline
Roqos        & 1.30.24          \\\hline
HomeHalo     & 1.0.0.8      \\\hline
Fingbox & 0.5-2ubuntu4 \\\bottomrule
\end{tabular}

\label{table:listdevices}
\end{table}

\begin{table}[H]
\centering
\caption{List of parental control \androidsols{}. 
\formatSpecial{*} denotes versions downloaded from vendor websites 
with extra features; \formatSpecial{P}, \formatSpecial{T} refer to premium and trial versions, respectively.
}
\begin{imageonly}\resizebox{1\columnwidth}{!}{%
\setlength\tabcolsep{2pt}
\begin{tabular}{p{0.55in}clc}
\textbf{Solution} & \textbf{Installs} & \textbf{App package name} & \textbf{Version} \\ \midrule
\meetcircle{} & 10K+ & {com.meetcircle.circle} & \makecell{P2.8.0.2} \\ \hline 
\familytime & 500K+ & \makecell[l]{io.familytime.dashboard\\ io.familytime.parentalcontrol\\ io.familytime.parentalcontrol(\formatSpecial{*})} & \makecell{P2.1.0.210\\ P3.0.5.3196.ps\\ P4.0.6.4209.web}\\ \hline
\famisafe{} & 500K+ & {com.wondershare.FamiSafe}& \makecell{P3.0.9.107}\\ \hline 
\findmykids{} & 10M+ &\makecell[l]{org.findmykids.app \\ org.findmykids.child} & \makecell{T1.9.9\\ T1.9.9}\\ \hline
\kidoz{} & 1M+ & \makecell[l]{com.kidoz \\ com.kidoz.demo.go(\formatSpecial{*})
} & \makecell{P4.0.5.8\\ P4.0.6.3} \\ \hline
\KidControl{} & 1M+ &\makecell[l]{ru.kidcontrol.gpstracker 
\\ app.gpsme 
} & \makecell{T4.0.9\\ Tk5.2.10}\\\hline
\kidsplace{} & 5M+ &\makecell[l]{
com.kiddoware.kidsplace \\ com.kiddoware.kidsafebrowser\\
com.kiddoware.kidsvideoplayer\\ com.kiddoware.kidsplace.remotecontrol\\ com.kiddoware.kidspictureviewer} &\makecell{ 
P3.5.6 \\ P1.7.8\\ 
P1.7.8 \\ P1.4.5\\ P1.0.9} \\\hline
\life{} & 50M+ & com.life360.Android.safetymapd & \makecell{T18.7.1} \\ \hline
\mmguadian{} & 1M+ &\makecell[l]{com.mmguardian.parentapp\\ com.mmguardian.childapp \\ 
com.mmguardian.childapp(\formatSpecial{*})
} 
&\makecell{P3.6.4\\ P3.7.7 
\\ P10003.9.5 } \\\hline
\mobilefence{} & 1M+ &\makecell[l]{com.mobilefence.family \\ com.mobilefence.family.plugin(\formatSpecial{*})
} & \makecell{T2.9.3.1 \\ T1.4}\\ \hline
\qustodio{} & 1M+ & \makecell[l]{
com.qustodio.qustodioapp \\ com.qustodio.qustodioapp(\formatSpecial{*})
}
 & \makecell{T180.14.2.2-family \\ T680.14.2.2-family } \\ \hline
\screentime{} & 1M+ &\makecell[l]{com.screentime.rc \\ com.screentime} & \makecell{
T3.11.23 \\ T5.3.23 }  \\ \hline
\secureteen{} & 1M+ &\makecell[l]{com.infoweise.parentalcontrol.secureteen \\ com.infoweise.parentalcontrol.secureteen.child \\ com.infoweise.parentalcontrol.secureteen.child(\formatSpecial{*})
} &\makecell{T8.0.0 \\ T1.6000.5 \\ T1.7001.0 } \\ \bottomrule
\end{tabular}}
\end{imageonly}
\label{table:listapp_tab}
\end{table}
\begin{table}[ht]
\centering
\caption{List of parental control \windowsapps{} and their corresponding websites' popularity (traffic/ranking); we analyzed the lasted versions available. 
}

\resizebox{0.9\columnwidth}{!}{%
\setlength\tabcolsep{1pt}
\begin{tabular}{l|c|c|c} 
\textbf{Application} & \textbf{\# of visits/day} & \textbf{Main countries} & \textbf{World Alexa rank} \\ \midrule
Qustodio & 27K & US & 85,673 \\
Kaspersky & 1,400K & IN, US, RU & 2,114 \\
Dr. Web & 84K & US & 40,515 \\
Norton & 6,400K & US & 431 \\
Spyrix & 21K & UK & 230,966 \\
Kidswatch & NA & NA & 2,175,932 \\
KidLogger & 4.2K & PE & 156,645 \\
Kurupira & 17K & BR & 84,918 \\ \bottomrule
\end{tabular}
}
\label{table:listofWindowsapps}
\end{table}

\begin{table}[H]
\caption{List of parental control Chrome extensions.}
\centering

\label{table:listext_tab}
\resizebox{\columnwidth}{!}{%
\setlength\tabcolsep{1pt}
\begin{tabular}{llll}
\textbf{Extension} & \textbf{Installs} & \textbf{ID} & \textbf{Version} \\ \midrule
Blocksi Web Filter & 40K+ & pgmjaihnmedpcdkjcgigocogcbffgkbn & 1.0.144 \\ \hline
Parental control & 3K+ & bdjgolepmhcchlgncgkmobepknekjbkd & 1.0.22 \\ \hline
TinyFilter & 20K+ & epniipcfpbjliciholgdeipceecgcfmj & 2016.11.1.1 \\ \hline
Porn Blocker 
& 10K+ & kmillccnmojidmkhhjngjlalnbhpobcl & 1.5. 2 \\ \hline
Adult Blocker
& 80K+ & onjjgbgnpbedmhbdoikhknhflbfkecjm & 6.2.8 \\ \hline
Anti-porn addon 
& 20K+ & peocghcbolghcodidjgkndgahnlaecfl & 2.20.0 \\ \hline
MateCode Blocker
& 80K+ & gppopmmjibhcboobpmfombbkoehgicoh & 1.0.5 \\ \hline
MetaCert 
& 20K+ & dpfbddcgbimoafpgmbbjiliegkfcjkmn & 0. 10.18 \\ \hline
FamilyFriendly & 7K+ & epdelmeadnnoadlcalkmacoopocdafnp & 0.9.0 \\ \hline
Kids Safe Web & 3K+ & lakceedfffnfheaipjadbcndkldlplnd & 1.0.7 \\ \bottomrule
\end{tabular}%
}
\end{table}

\setcounter{footnote}{16}

\subsection{Techniques Adopted by \androidsols{}} 
Table \ref{table:moniteringtechniques} provides a summary of some techniques adopted by the  analyzed \androidsols{}. Four \androidsols{} (MMGuardian, MobileFence, Qustodio, and SecureTeen),  distributed via their company websites, support additional features compared to their Google Play store version.\footnotetext{\famisafe{} \androidapp{} gets full access to the child's YouTube account including rights to view, edit, delete the child's YouTube videos and playlists, and rate videos, post, edit/delete comments and captions.} 

\begin{table}[H]
\centering
\caption[]{Techniques used to monitor child activities including web filtering, phone calls, SMS, and social media. \newline
\CIRCLE: refers to service supported by Google Play version;  \newline\LEFTcircle: refers to a feature supported by a version distributed via the company website.}
\label{table:moniteringtechniques}
\begin{tabular}{@{}l}
\includegraphics[]{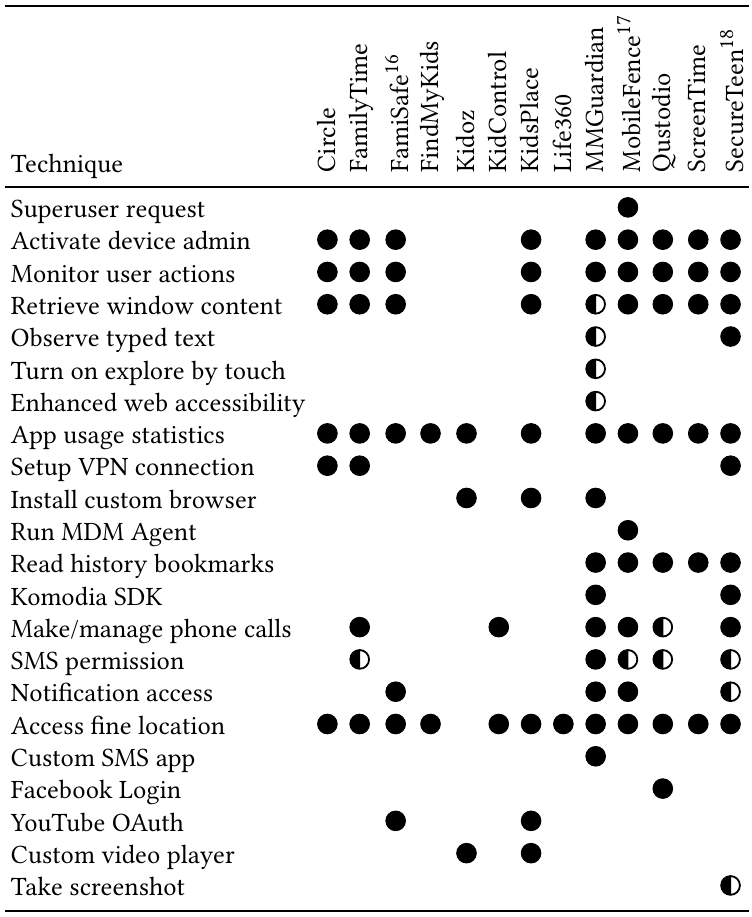}
\end{tabular}%%%\resizebox{0.9\columnwidth{}}{!}{%
\end{table}

\setcounter{footnote}{17}

\subsection{Third-Parties Analysis Results} 
\begin{figure}[H]
\centering
\includegraphics[width=\linewidth, trim={0.2cm 0.41cm 0.4cm 0.0cm}, clip]{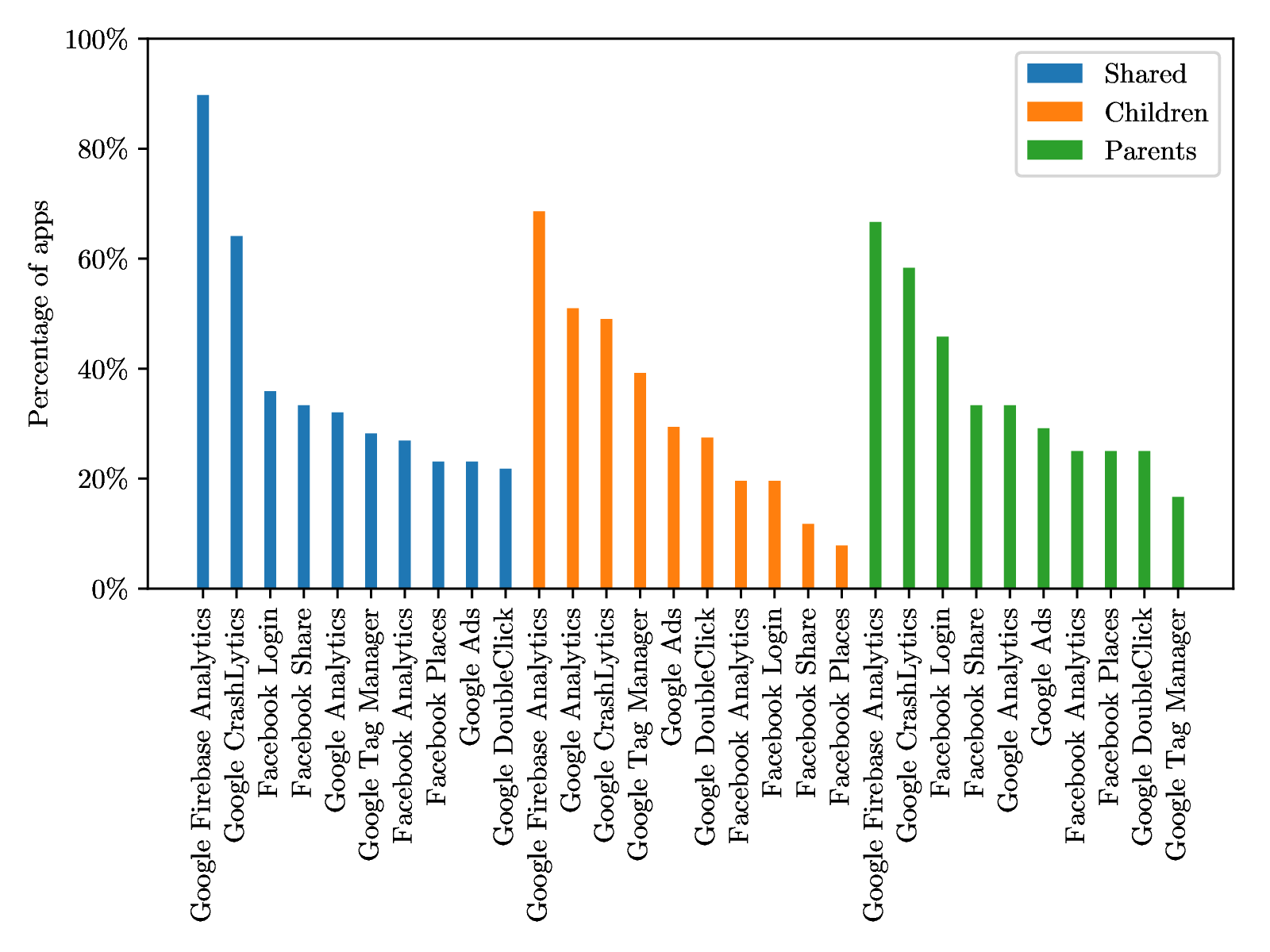}
\caption{Tracking SDKs present in \androidapps{} found through static analysis, see Sec.~\ref{Sec:3rdparties}.}

\label{fig:Percstaticanalysis}
\end{figure}

\begin{figure}[H]
\centering
\includegraphics[width=\linewidth, trim={0.2cm 0.4cm 0.39cm 0.0cm}, clip ]{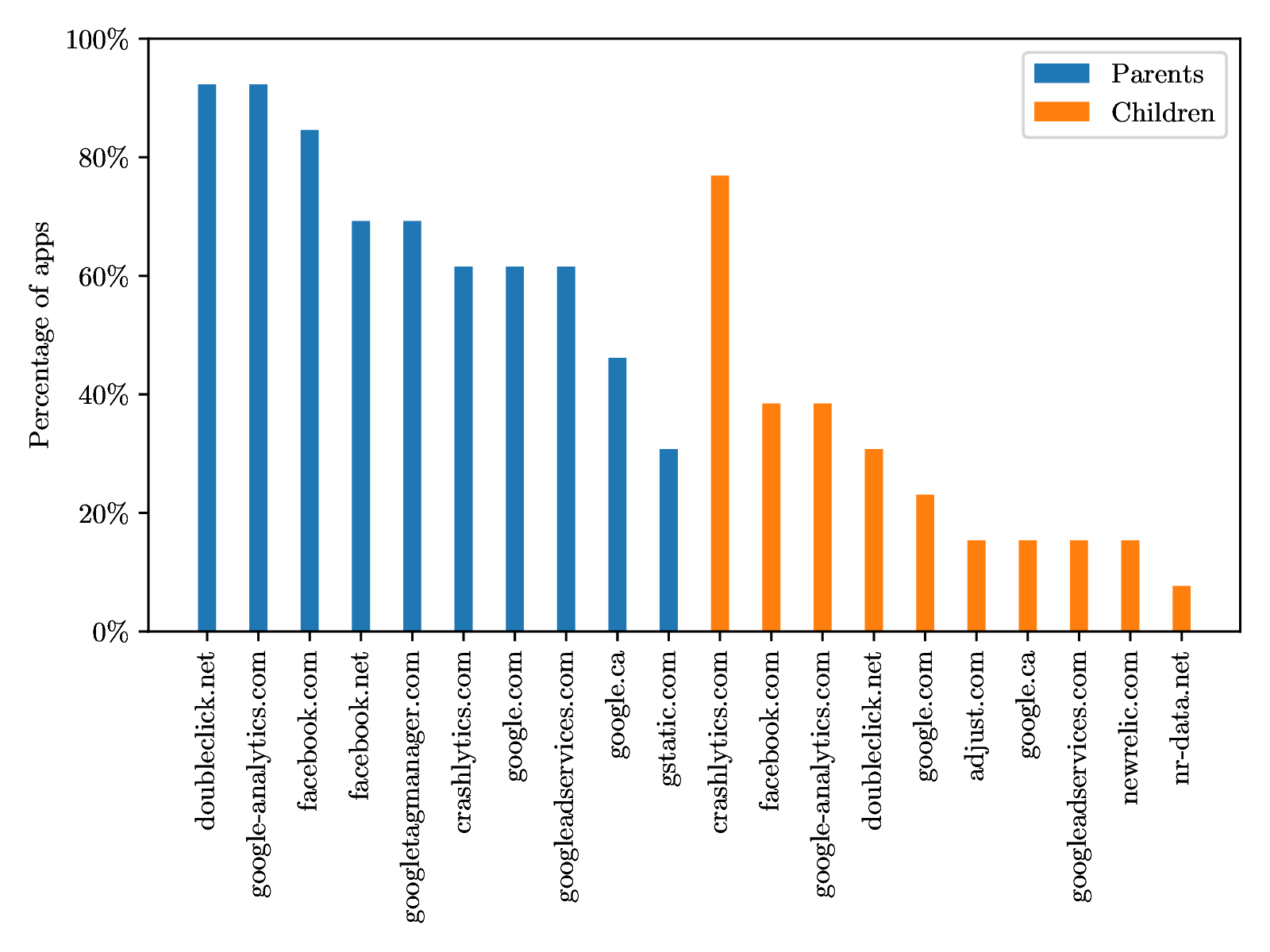}
\caption{Tracking SDKs present in \androidsols{} found through dynamic analysis, see Sec.~\ref{Sec:3rdparties}.}
\label{fig:Percdynamicanalysis}

\end{figure}

Table~\ref{table:countoftrackingsdks} shows the use of third-party\setcounter{footnote}{17}\footnotetext{{\mobilefence{} initially setup by default to monitor both the child and parent devices.}}\setcounter{footnote}{18}\footnotetext{\secureteen{} \androidapp{} uses a keylogger to record all social media activities on the child device.} tracking \SDKs{} in the analyzed \totalevaluated{} \androidapps{}. We used MOBSF~\cite{MOBSF} to extract the list of third-party tracking \SDKs{} from all \app{}s based on Exodus-Privacy's tracker list.  
On average, we found 4.5 \SDKs{} per \app{} (max 10 \SDKs{}) in \purechildrenapps{}{}. The average number of \SDKs{} increases to about \ 5.3 \SDKs{} per \app{} in \mixedchildparentenapps{} and \parentapps{}{}.  We also found Google Firebase Analytics, Google CrashLytics are present in over 50\% of all types of \app{}s; see Fig.~\ref{fig:Percstaticanalysis}. We also identified tracking third-party \domains{} from network traffic generated during our dynamic analysis; see Fig.~\ref{fig:Percdynamicanalysis}.

\begin{table}[H]
\centering
\caption{Use of tracking \SDKs{} in \purechildrenapps{}, \mixedchildparentenapps{} (i.e., the same is used by both parents and children), and \parentapps{} found through static analysis.}
\label{table:countoftrackingsdks}
\begin{imageonly}
\begin{tabular}{lccc}
\toprule
\textbf{} & \textbf{\HPurechildrenapps{}} & \textbf{\HMixedchildparentenapps{}
} & \textbf{\HParentapps{}} \\\midrule
\# \androidapps{} & 51 & 78 & 24 \\
\# Unique tracking \SDKs{} & 35 & 41 & 31 \\
\# \app{}s with tracking \SDKs{} & 44 & 73 & 22 \\\midrule
Average \# \SDKs{} per \app{} & 4.5 & 5.3 & 5.4 \\
Max \# \SDKs{} per \app{} & 10 & 22 & 12 \\ \bottomrule
\end{tabular}
\end{imageonly}
%}
\end{table}

\subsection{Data Sharing and Privacy Leaks} 
Table~\ref{table:list_pii} lists the personal information used to detect PII data in network traffic. Tables~\ref{table:leakviahttp} and \ref{table:sharethirdparty} show the PII transmitted by \androidsols{} in plaintext through HTTP, and PII shared with third-parties, respectively.

\begin{table}[H]
\centering
\caption{The list of personal information used to detect PII data in network traffic.}
\label{table:list_pii}
\begin{imageonly}
\resizebox{0.93\columnwidth}{!}{%
\begin{tabular}{ll}
\toprule
\textbf{PII} & \textbf{Description} \\ \midrule
AAID & Android Advertising ID \\
Android ID & Android ID generated on device setup \\
GSF ID & Google Services Framework ID \\
Phone Serial & Mobile serial number \\
IMEI & Phone equipment ID \\
SIM ID & SIM card ID \\
AP BSSID & MAC addresses of used hotspots \\
AP SSID & SSIDs of used hotspots \\
Nearby AP BSSID & MAC addresses of surrounding hotspots \\
Nearby AP SSID & SSIDs of surrounding hotspots \\
MAC Address & MAC address of the WiFi interface \\
IP address & IP address of the WiFi interface \\
BD ADDR & MAC address of the Bluetooth interface \\
Google Email & Google play account email address \\
User credentials & Account ID and password \\
Name & User's first and last names \\
Email & User's email address \\
Phone \# & User's phone number \\
Geolocation & Latitude \& Longitude  \\
Contacts & Contact list entries \\
Browsing history & Visited URLs in browser \\
Used App & Apps used on the device \\
Installed Apps & Apps installed on the device \\
Social messages & SMS/social media messages \\ 
Search history  & Search strings used on Google or Youtube \\
Mobile carrier & User's mobile carrier \\
Address &  {User's address (street name, city, country,   and postal code)}\\
\bottomrule
\end{tabular}
}
\end{imageonly}
\end{table}

\begin{table}[H]
\centering
\caption{\androidsols{} sending sensitive data in plaintext.}
\label{table:leakviahttp}
\begin{imageonly}
\resizebox{0.93\columnwidth{}}{!}{%
\begin{tabular}{lll}
\toprule
\textbf{Solution} & \textbf{Data} & \textbf{Destination} \\ \midrule
\kidoz{} & Account username/password & \url{kidoz.net} \\
\kidoz{} & Child name & \url{kidoz.net} \\
\kidsplace{} & Child Android ID & \url{kiddoware.com} \\
\kidsplace{} & Child phone serial & \url{kiddoware.com} \\
\kidsplace{} & Parent email & \url{kiddoware.com} \\
\KidControl{}{} & Child Geolocation & \url{kid-control.com} \\
\KidControl{}{} & Parent email  & \url{kid-control.com} \\
\KidControl{}{} & Parent name & \url{kid-control.com} \\
\life{} & Child name  & \url{pubnub.com} \\
\life{}{} & Parent name & \url{pubnub.com} \\
\mmguadian{} & Account username/password~\formatSpecial{*} & \url{mmguardian.com} \\
\mmguadian{} & Parent email & \url{mmguardian.com} \\
\mmguadian{} & Parent IMEI  & \url{mmguardian.com} \\
\mmguadian{} & Parent/child phone \# & \url{mmguardian.com} \\
\mmguadian{} & Parent phone serial & \url{mmguardian.com} \\
\mmguadian{} & Browsing history & \url{komodia.com} \\
\mmguadian{} & Parent AAID & \url{mmguardian.com} \\
\mmguadian{} & Child Geolocation  & \url{mmguardian.com} \\
\mmguadian{} & Child installed apps & \url{mmguardian.com} \\

\secureteen{} & Parent email & \url{secureteen.com} \\
\secureteen{} & Parent name & \url{secureteen.com} \\
\bottomrule
\multicolumn{3}{l}{
\makecell[lt]{\formatSpecial{*}: The parent's password is hashed using SHA1 without salting.}
 }
\end{tabular}%
}
\end{imageonly}
\end{table}

\begin{table}[H]
\caption{Sharing PII with third-parties.}
\centering
\begin{imageonly}
\resizebox{\columnwidth}{!}{
\setlength\tabcolsep{1pt}
\begin{tabular}{lll}
\toprule
\textbf{Solution} 			& \textbf{Shared PII} & \textbf{3rd-parties (number, domains [max.\ 2])}~\formatSpecial{*} \\ \midrule

\meetcircle{} 			& Child Android ID & 1 (\url{kochava.com})\\

\meetcircle{} 			& Parent Android ID & 2 (\url{kochava.com},~\url{mixpanel.com})\\

\meetcircle{} 			& Child/parent AAID & 1 (\url{kochava.com})\\

\meetcircle{} 			&  Child/parent AP BSSID & 1 (\url{kochava.com})\\

\meetcircle{} 			&  Child/parent AP SSID & 1 (\url{kochava.com})\\

\meetcircle{} 			& Child name & 1 (\url{intercom.com})\\

\meetcircle{} 			&  Parent email & 5 (\url{intercom.com},~\url{apptentive.com})\\

\meetcircle{} 			&   Parent name & 3 (\url{facebook.com}, ~\url{mixpanel.com})\\

\meetcircle{} 			& Child mobile carrier & 1  (\url{kochava.com}) \\
\meetcircle{} 			& Parent mobile carrier & 2  (\url{kochava.com},~\url{apptentive.com}) \\

\familytime{} 			&  Child name & 1 (\url{facebook.com}) \\
\familytime{} 			&  Child email & 1 (\url{facebook.com}) \\
\familytime{} 			&  Child phone \# & 1 (\url{facebook.com}) \\

\familytime{} 			&  Parent email & 11  (\url{doubleclick.net},~\url{facebook.com}) \\

\familytime{} 			& Parent name & 11  (\url{fastspring.com},~\url{google-analytics.com}) \\

\familytime{} 			& Parent address & 1  (\url{fastspring.com}) \\

\familytime{} 			& Parent AAID & 1  (\url{facebook.com}) \\

\familytime{} 			& Parent mobile carrier & 1  (\url{facebook.com}) \\

\familytime{} 			& Parent phone \# & 1 (\url{fastspring.com}) \\

\famisafe{} & Child AAID & 1  (\url{graph.facebook.com}) \\
\famisafe{} 			&  Child name & 1 (\url{facebook.com}) \\

\famisafe{} 			&  Child Geolocation & 1 (\url{maps.googleapis.com}) \\

\famisafe{} 			&  Child browsing history & 2 (\url{facebook.com}, \url{google-analytics.com}) \\
\famisafe{} &  Child device carrier & 1  (\url{graph.facebook.com}) \\
\kidoz{} 			& Child AAID & 1 (\url{googleapis.com}) \\

\findmykids{} 			& Child/parent AAID & 3 (\url{yandex.net}, ~\url{facebook.com}) \\

\findmykids{} 			&  Child/parent Android ID & 1 (\url{yandex.net})\\

\findmykids{} 			&  Child Geolocation & 2 (\url{openstreetmap.org},~\url{yandex.net}) \\

\findmykids{} 	& Child Nearby AP BSSID & 1 (\url{yandex.net})\\

\findmykids{} 	 & Child Nearby AP SSID & 1 (\url{yandex.net})\\

\findmykids{} 			&  Child/parent mobile carrier & 1 (\url{facebook.com}) \\

\KidControl{} 			& Child Geolocation & 1 (\url{openstreetmap.org}) \\

\KidControl{} 			& Parent email & 1 (\url{firestore.googleapis.com}) \\

\kidsplace{} 			& Child AAID & 2 (\url{google-analytics.com},~\url{onesignal.com}) \\

\kidsplace{} 			& Child mobile carrier & 1 (\url{onesignal.com}) \\

\kidsplace{} 			& Child Geolocation & 1 (\url{maps.googleapis.com}) \\

\kidsplace{} 			& Parent email & 1 (\url{sendgrid.com}) \\

\life{} 			& Child Android ID & 1 (\url{branch.io}) \\

\life{} 			& Parent Android ID & 2 (\url{branch.io},~\url{amazonaws.com}) \\

\life{} 			& Child AAID & 3  (\url{appsflyer.com},~\url{branch.io}) \\
\life{} 			& Parent AAID & 4  (\url{appsflyer.com},~\url{facebook.com}) \\
\life{} 			& Child/parent name & 2  (\url{braze.com},~\url{pubnub.com}) \\
\life{} 			& Child email & 2 (\url{helpshift.com},~\url{braze.com}) \\

\life{} 			& Parent email & 3  (\url{helpshift.com},~\url{braze.com}) \\ 

\life{} 			& Child/parent local IP & 1 (\url{branch.io}) \\

\life{} &  Child/parent Geolocation & 3 (\url{locationiq.com},~\url{braze.com})\\

\life{} 			&  Parent phone \# & 1 (\url{amazonaws.com}) \\

\life{}			& Parent AP BSSID & 1 (\url{amazonaws.com})\\

\life{}			&  Parent AP SSID & 1 (\url{amazonaws.com})\\

\life{} 			& Child/parent mobile carrier & 3  (\url{appsflyer.com},~\url{braze.com}) \\

\mmguadian{} 			& Child/parent AAID & 2  (\url{facebook.com},~\url{googleadservices.com}) \\
\mmguadian{} 			& Child browsing history & 1 (\url{komodia.com}) \\
\mmguadian{} 			& Child/parent mobile carrier & 1 (\url{facebook.com}) \\

\mobilefence{} 			& Parent email \& name & 1 (\url{livechatinc.com}) \\

\mobilefence{} 			&Parent AAID & 1 (\url{googleadservices.com}) \\
 
\mobilefence{} 			& Child Geolocation & 2 (\url{googleapis.com},~\url{amazonaws.com}) \\

\mobilefence{} 			& Child browsing history & 1 (\url{google.com}) \\

\qustodio{} 			& Child/parent Android ID & 2 (\url{amazonaws.com},~\url{rollout.io})

\\
\qustodio{} 			& Child/parent AAID & 1 (\url{adjust.com})\\

\qustodio{} 			& Parent  email & 3 (\url{adroll.com},~\url{braze.eu})\\
\qustodio{} & Parent  name & 1 (\url{referralcandy.com})\\

\qustodio{} 			& Child used app & 1 (\url{google-analytics.com}) \\

\qustodio{} 			& Child/parent mobile carrier & 1  (\url{braze.eu}) \\
\screentime{} 			&  Child/parent AAID & 1 (\url{graph.facebook.com}) \\

\screentime{} 			& Child Android ID & 4 (\url{facebook.com},~\url{googleapis.com})\\
\screentime{} 			& Parent Android ID & 1 (\url{appspot.com})\\

\screentime{} 			&  Child name & 3 (\url{appspot.com},~\url{facebook.com}) \\

\screentime{} 			&  Parent email \& name & 2 (\url{appspot.com},~\url{facebook.com}) \\
\screentime{} 			&   Child 
Geolocation & 4 (\url{google.com},~\url{googleapis.com}) \\

\screentime{} 			& Child installed apps & 1 (\url{appspot.com}) \\
\screentime{} 			&  Parent Mobile carrier & 1 (\url{facebook.com}) \\
\screentime{} 			& Child/parent mobile carrier & 1  (\url{graph.facebook.com}) \\

\secureteen{} 			&  Parent email & 27 (\url{adroll.com},~\url{ads.yahoo.com}) \\

\secureteen{} 			& Child browsing history & 1 (\url{komodia.com}) \\

\secureteen{} 			& Child Geolocation & 1 (\url{google.com}) \\ 
\bottomrule
\multicolumn{3}{l}{
\makecell[lt]{
 \formatSpecial{*}: Number of domains limited to 2 to fit display; \formatSpecial{AAID} refers to Android Advertising ID; 
  }
 }\\
\multicolumn{3}{l}{
\makecell[lt]{
We use the word \quotesd{domain} to refer to second-level domains.}}
\end{tabular}}
\end{imageonly}
\label{table:sharethirdparty}
\end{table}

\end{document}